\documentclass[aps,prd,reprint,superscriptaddress,nofootinbib]{revtex4-2}

\usepackage{ifxetex}
\ifxetex
   \usepackage{polyglossia}
   \setdefaultlanguage{english}
   \usepackage{fontspec}
\else
   \usepackage[utf8]{inputenc}
\fi
\usepackage{amsmath}
\usepackage{graphicx,color,overpic,mathtools}
\usepackage{amsthm}
\usepackage{bm}
\usepackage{amsmath}
\usepackage{upgreek}
\usepackage{amssymb}
\usepackage{setspace}
\usepackage{hyperref}
\usepackage{xcolor}
\PassOptionsToPackage{normalem}{ulem}
\usepackage{ulem}
\usepackage{tensor}

\usepackage{graphicx,color}
\usepackage{bbold}
\usepackage{overpic}
\usepackage{hyperref}
\usepackage{pgf}
\usepackage{tikz}

\newcommand{\noprint}[1]{}

\newcommand{\tr}{{\text{tr}}}

\newcommand{\ii}{\mathrm{i}}
\renewcommand{\d}{\mathrm{d}}
\renewcommand{\Re}{\mathrm{Re}}
\renewcommand{\Im}{\mathrm{Im}}
\newcommand{\nn}{\nonumber}

\usepackage{braket}
\usepackage{bbm}

\providecommand{\norm}[1]{\lVert#1\rVert}

\usepackage{braket}
\usepackage{bbm}

\providecommand{\norm}[1]{\lVert#1\rVert}

\begin{document}
\title{Causality and signalling in non-compact detector-field interactions}

\author{Jos\'e de Ram\'on}
\affiliation{Institute for Quantum Computing, University of Waterloo, Waterloo, Ontario, N2L 3G1, Canada}
\affiliation{Department of Applied Mathematics, University of Waterloo, Waterloo, Ontario, N2L 3G1, Canada}

\author{Maria Papageorgiou}
\affiliation{Institute for Quantum Computing, University of Waterloo, Waterloo, Ontario, N2L 3G1, Canada}
\affiliation{Department of Applied Mathematics, University of Waterloo, Waterloo, Ontario, N2L 3G1, Canada}
\affiliation{Division of Theoretical and Mathematical Physics, Department of Physics, University of Patras, 26504, Patras, Greece}

\author{Eduardo Mart\'in-Mart\'inez}
\affiliation{Institute for Quantum Computing, University of Waterloo, Waterloo, Ontario, N2L 3G1, Canada}
\affiliation{Department of Applied Mathematics, University of Waterloo, Waterloo, Ontario, N2L 3G1, Canada}
\affiliation{Perimeter Institute for Theoretical Physics, 31 Caroline St N, Waterloo, Ontario, N2L 2Y5, Canada}


\begin{abstract}
In this paper we analyze the problem of ``apparent'' superluminal signalling and retrocausation that can appear for particle detector models when considering non-compactly supported field-detector interactions in quantum field theory in curved spacetimes and in relativistic quantum information protocols. For this purpose, we define a signalling estimator based on an adapted version of the quantum Fisher information to perturbative regimes. This allows us to study how the internal dynamics of the detectors (for example the gap between the detector energy levels) have an impact on the ability of a particle detectors to communicate with one another. Moreover, we show that, very generally, even for detectors with infinite tails in space and time, if the tails decay exponentially, one can define an effective lightcone, outside of which signalling is negligible. This provides concrete evidence supporting the use of non-compact (but exponentially decaying) detector smearings in protocols of relativistic quantum information.

\end{abstract}
\maketitle

\section{Introduction}

Particle detector models, such as the Unruh-DeWitt (UDW) detector~\cite{PhysRevD.14.870,DeWitts}, can be thought of as a framework to model the measurement of quantum fields. As their name suggests, these models were originally intended to tackle the challenges associated to the notion of \textit{particle} in relativistic quantum field theory (QFT), that is, to deal with  the complications that a naive notion of particle introduces in scenarios involving non-inertial observers or curved spacetimes. Particle detectors have provided physical insight in a wide range of scenarios, including black hole evaporation \cite{PhysRevD.15.2738} and the entanglement structure of QFT \cite{VALENTINI1991321,Reznik2003,PhysRevD.94.064074} to more mundane descriptions of the light-matter interaction in quantum optics experiments and quantum communication \cite{PhysRevLett.114.110505}.

The simplest formulation of a particle detector model is perhaps the pointlike Unruh-DeWitt model. In this model the localization of the detector is considered to be constrained to a particular wordline. The practice of smearing the interaction of the detector with the field in spacetime became popular in the particle detector literature as a way to ameliorate the UV behaviour of detector transition rates~\cite{Louko2006HowOD, Sebastian_Schlicht_2004}. Later on, it was shown that smeared detector-field interactions 
can better capture the physics of light-matter interaction \cite{PhysRevD.97.105026,PhysRevD.94.064074,PhysRevA.103.013703}, or for example, the physics of superconducting qubits \cite{PhysRevA.96.052325}. Despite these advantages, smeared couplings are not devoid of their own fundamental issues. In particular, the non-pointlike (extended) detector-field interaction can introduce friction with relativistic causality and enable superluminal signaling in some regimes~\cite{PhysRevD.92.104019,PhysRevD.103.085002}.

As reported in \cite{PhysRevD.103.085002}, detector-field interactions that are strictly bounded in space-time (compactly smeared) do not present problems with superluminal signalling in bipartite detector scenarios (both perturbatively and non-perturbatively). The reason (see \cite{PhysRevD.103.085002} for further details) is that if the interactions between each of the detectors and the field are causally orderable (i.e., the regions can be separated by a Cauchy surface) then the dynamics factorizes in a way that prevents faster than light signalling or retrocausation. It was shown there that, therefore, the problem of superluminal signalling facilitated by the possibly non-relativistic description of the detector models at hand does not represent a problem in bipartite situations. Furthermore, it was also shown that in multipartite scenarios (scenarios with many detectors), there is no incompatibility with causality to leading order in perturbation theory \cite{PhysRevD.103.085002} (see also \cite{Benincasa_2014}).


However, in situations where the interactions are not compactly supported (and, as a consequence, not causally orderable), rigorously speaking, non-compact particle detectors are always in some degree of causal contact no matter how far away they are. Of course we know that in physical scenarios (like modeling atomic probes interacting with the electromagnetic field) the tails of the spacetime localization of the atoms can be exponentially suppressed, so effectively there is perhaps a regime where two atoms interacting for a finite amount of time and sufficiently far away will be  \textit{effectively} causally disconnected. Moreover, it is common to consider toy models where one abandons the compact support of the interaction region in order to achieve analytical results, e.g. it is common to make use of Gaussian profiles for both smearing and switching functions (adiabatic switchings) whose impact on signalling has been studied to some extent in \cite{PhysRevD.92.104019}. 


In this paper we are going go beyond the results in \cite{PhysRevD.92.104019} by building a signalling estimator that can characterize what are the regimes where one can use non-compact smearings and still be able to consider that two far apart detectors are, for all intents and purposes, causally disconnected. For this we will use the theory of optimal parameter estimation to characterize the influence of a particle detector localized to a regions of spacetime on an other detector in a different region. By doing so we will be able to fully characterize the dependence on all the parameters of the detectors and show that even the internal energy gap of the detectors can play a fundamental role in deciding whether to smeared detectors are or not in effective spatial separation, something that previous literature did not cover.

This work is divided in the following sections.
In section \ref{sec:zoo} we review a selection of different UDW-type detector models employed in the literature, with the purpose of clarifying their differences and similarities. Further, we establish a unifying language to analyze the causality issues of general detector models in general globally hyperbolic space-times, and why non-compact smearings are sometimes introduced. In section \ref{sec:estimator}, we build towards a quantitative analysis of signalling in general detector models. For weak couplings, using peturbation theory, we show in which cases the microcausality condition for the quantum field prevents the signaling between non-compact detectors that are strongly localized in regions in spacelike separation, generalizing the result in~\cite{PhysRevD.92.104019}. In the same section, we will define a new signalling estimator, based on the quantum Fisher information, that can be used to quantify the impact that the `overlap' of the tails of non-compact detectors influence whether we can consider the detectors effectively spacelike separated.  In section \ref{sec:examples} we provide formal expressions for this signalling estimator for some of the models considered in section \ref{sec:zoo}. In section \ref{betweentwolevels} we perform an in-depth analysis of the signaling estimator for the simple, yet relevant case of UDW detectors interacting through a massless scalar field in flat spacetime in 3+1 dimensions. Finally, in section \ref{sec:bounds} we provide exponential bounds for signalling between UDW detectors in flat spacetimes based solely on analyticity properties. We conclude and summarize in section \ref{sec:conclusions}.

\section{Particle detector models and non-compact smearings}\label{sec:zoo}

In this section we will review some of the most relevant variations of particle detector models that have been used in previous literature. Our goal is to establish a unifying language that allows us to capture the relevant features of each model and study their causal behaviour. In particular, we will review and motivate the different ways in which the detector-field interactions can make use of non-compact spacetime smearings. 

Each detector model is prescribed by an interaction Hamiltonian density that models when, where and how the detector is coupled to the field. The locality properties of each interaction Hamiltonian will encode the causal behavior of each model. For simplicity, we will assume that the detector couples to a real scalar field in all cases\footnote{There is a wealth of literature that establishes in what regimes a scalar field coupling captures the relevant features of the light matter interaction~\cite{PhysRevD.94.064074,PhysRevA.103.013703,  PhysRevD.87.064038} or can model high-energy processes involving spinor fields~\cite{PhysRevD.93.024019, PhysRevD.104.105021}.}.

 Perhaps the best-known detector model that has been considered in the literature is the Unruh-DeWitt (UDW) detector model. In its most popular version, the localization of the detector is considered to be constrained to a time-like trajectory. This is prescribed by the interaction Hamiltonian that generates translations with respect to the proper time $\tau$ associated to the detector's trajectory. In the interaction picture, this Hamiltonian is given by
 \begin{equation}\label{UDWpoint}
    \hat H_{\text{pt}}=\lambda\chi(\tau)\hat D(\tau) \otimes\hat\phi(\mathsf{x}(\tau)).
\end{equation}
 Here   $\lambda$ is the coupling strength, $\chi (\tau)$ is the switching function, which is usually assumed to be  integrable, and $\mathsf{x}(\tau)$ is the spacetime trajectory of the detector parametrized by its proper time $\tau$.   The Hamiltonian couples the field along the wordline of the detector to an internal degree of freedom of the detector $\hat{D}$.

The point-like model has been proven to not have any problems with the covariance of its predictions, as well as being completely causal~\cite{PhysRevD.92.104019,2020broken,PhysRevD.103.085002}.  Despite this, the pointlike model can exhibit ultraviolet divergences related to the coincidence limit of the time-ordered $n$-point functions. For the purpose of regularization, one can consider smearing the detector with a family of test-functions (e.g. Gaussian or Lorentzian functions parametrised by their width ${R}$)  to properly define the point-like limit as (${R} \rightarrow 0$). 

The simplest generalisation of the pointlike interaction Hamiltonian \eqref{UDWpoint} involves a spatially smeared field operator which captures the spatial extension of the interaction
\begin{align}
   \hat H_{\text{sm}} =\lambda \chi(t) \hat{D}(t)\otimes \int \d^n \bm x\, F(\bm x)\hat{\phi}(t,\bm x) \label{h1}.
\end{align}
 Here $F(\bm x)$ determines the interaction's spatial profile in the $(t,\bm x)$. If the smearing and switching functions are compactly supported over a spacetime region $\mathcal{O}$, the interaction can be thought of as happening only over that spacetime region, i.e., $\mathcal{O}= \text{supp}\left[ F(\bm x) \chi(t)\right]$ which we will call the interaction region. 
 The spatial extension of the interaction region can be typically motivated on physical  grounds. From a purely phenomenological point of view, we advocate for the most conservative interpretation of~\eqref{h1},  the smearing just carries information about where in space the interaction occurs and does not `belong' either to the detector or the field, as it was also pointed out in \cite{PhysRevA.96.052325}. The interaction region can be significantly larger than the actual size of the detector, as it happens in superconducting circuits where the detector (a superconducting flux or transmon qubit) is of macroscopic  size \cite{PhysRevA.96.052325}.
 
 For calculational convenience, it is common to choose non-compact smearing and switching functions. When one does this, strictly speaking, the interaction is not localised and analyzing the causal behavior of the model becomes more subtle \cite{PhysRevD.92.104019}. 
 


Regarding the choice of internal degree of freedom of the detector, a common choice is to consider that the detector is a two-level quantum system with an energy gap $\Omega$. However, the model is not restricted to this case and the moniker `Unruh-DeWitt detector' is nowadays used for most linear couplings of the form \eqref{h1}. For instance, another common detector model is the harmonic-oscillator Unruh-DeWitt variant (e.g. \cite{PhysRevD.101.125005})
\begin{align}
  \hat{H}_{\textsc{ho}}= \lambda \chi(t) \hat{\bm x}(t)\otimes \int \d^n \bm{x} F(\bm x) \,   \hat{\phi}(t,\bm x),  \label{h2}
\end{align}
where we substituted $\hat{D}(t)=\hat{x}(t)$ the position operator (or any other quadrature) of a harmonic oscillator written in the interaction picture. Using the terminology of Von Neumann-like measurements, the conjugate of $\hat{x}(t)$ can be interpreted as a continuous pointer variable that is linearly coupled to the field amplitude\footnote{ In the pointer variable analysis it is typical to assume no internal dynamics, i.e., a gapless continuous pointer variable (see example in subsection \ref{point}).}. 


An interaction Hamiltonian that is particularly designed for coupling the position operator of the detector `centre of mass' to the quantum field was introduced by Unruh and Wald in~\cite{PhysRevD.29.1047} where the field interact with the detector at the spatial points in the spectrum of the detector's position operator\footnote{An extension of the Unruh-Wald model that includes a coupling to a spin-like internal degree of freedom can be found here \cite{PhysRevD.101.036007}.}. The interaction Hamiltonian in this case is
\begin{align}
 \hat{H}_{\textsc{uw}}= \lambda \chi(t)\int \d^n \bm{x} \, \hat{\phi}(t,\bm x) \otimes \delta (\bm{x}-\hat{\bm x}_t).   \label{h3}
\end{align}
where we shortened notation $\hat{\bm x_t}\equiv \hat{\bm{x}}(t)$ for the interaction picture operators. The Dirac delta distribution $\delta (\bm{x}-\hat{\bm x}_t)$ is defined over an (interaction picture) orthonormal basis $\{\ket{i_t}\}$ of the Hilbert space of the detector:
\begin{align}
  \delta(\bm{x}-\hat{\bm x}_t)&\coloneqq \sum_{ij} \langle i_t | \delta (\bm{x}-\hat{\bm x}_t)| j_t \rangle \ket{i_t}\! \bra{j_t} \nonumber \\
  &=  \sum_{ij} \int  \d^n\bm{x}' \langle i_t | \delta (\bm{x}-\hat{\bm x}_t) \ket{\bm{x}'_t}\bra{\bm{x}'_t} j_t \rangle \ket{i_t}\!\bra{j_t} \nonumber \\
  &= \sum_{ij}\psi_i^*(\bm{x})\psi_j(\bm{x}) \ket{i_t}\!\bra{j_t} \label{delta}
\end{align}
If we define $F_{ij}(\bm{x})=\psi_i^*(\bm{x})\psi_j(\bm{x})$ the interaction Hamiltonian \eqref{h2} becomes
\begin{align}
   \hat{H}_{\textsc{uw}}=  \lambda \chi(t) \sum_{ij} \int \d^n \bm{x} F_{ij}(\bm{x})\hat{\phi}(t,\bm x) \otimes \ket{i_t}\bra{j_t} \label{h3b}
\end{align}
which consists of multiple terms of smeared field operators coupled to rank-one detector operators that correspond to all possible transitions between the elements of the basis $\{\ket{i_t}\}$, which we can understand as the eigenstates of a chosen observable.

Note that the smearing function of the field operator is not introduced by hand in \eqref{h3}, in contrast with \eqref{h2}. The smearing functions $F_{ij}(x)$ are not a freedom of the model, and instead they are determined by the kind of transition we are interested in. 

In fact, inspired in light-matter interaction models (see e.g.~\cite{PhysRevD.94.064074,PhysRevA.103.013703}), we can suggest yet a different interaction  Hamiltonian  designed to capture the features of the electromagnetic dipole coupling of a hydrogen-like atom with the electromagnetic field. The dipole interaction Hamiltonian of an electron in an atom and the electric field is of the form $\bm x\cdot\bm E(t,\bm x)$. We can write a scalar version of this coupling that still captures the same basic physics, by considering the coupling to a component of the electric field.  If we introduce a polarisation vector $\bm{\epsilon}$ we can write a dipole-like Hamiltonian like
\begin{align}
     \hat{H}_{\textsc{dp}}&=\lambda \chi(t)\int \d^n \bm{x} \, \bm{\epsilon} \cdot \bm{x}\, \hat{\phi}(t,\bm x) \otimes \delta (\bm{x}-\hat{x}_t)
\end{align}
If we bring this interaction Hamiltonian to the form \eqref{h3b} the smearing functions become $F_{ij}= \bm{\epsilon}\cdot\bm{x}\, \psi_i^*(\bm{x})\psi_j(\bm{x})$. A careful treatment of the modeling of light-matter interaction with UDW-type detectors beyond the scalar approximation can be found here \cite{PhysRevA.103.013703}.

More generally, one can consider couplings of the form 
\begin{align}
    \hat{H}_{\text{curr}}= \int \d^n \bm{x}\, \hat{J}(t,\bm x)\otimes \hat{\mathcal{O}}(t,\bm x) \label{anastop}
\end{align}
as it was proposed e.g. in \cite{ANASTOPOULOS2023169239}, 
where \mbox{$\hat{J}(t,\bm x)=e^{i \hat{p}\bm{x}}\hat{J}(t)e^{-i \hat{p}\bm{x}}$} is a general current operator for the detector system. There is a variety of quantum mechanical current operators in the literature, whose expectation values are thought of as defining the current of a moving charged quantum-mechanical particle, e.g., \cite{PhysRevA.86.012111}. In principle, the current operator can be derived from fundamental interactions in an effective field theory approach (e.g. in the case of neutrino detection \cite{PhysRevD.102.093003,PhysRevA.86.012111}). $\hat{J}(t)$ is any detector observable and $\hat{\mathcal{O}}(t,\bm x)$ is any field operator\footnote{Here we will focus on the case of linear coupling to the field amplitude. The treatment of non-linear couplings has been studied in the past, for instance to tackle particle detection protocols with complex scalar fields as well as well as with fermionic fields \cite{Takagi,PhysRevD.93.024019}, and to model photodetection protocols  with the Glauber model \cite{PhysRev.130.2529}} in their respective interaction pictures.  In this formalism the smearing function is motivated by the introduction of cross-grained variables for the detector system. Also, there is typically no switching function.

  Note that the Unruh-Wald model can be considered  a special case of  \eqref{anastop} for $\hat{J}(t)=\delta(\hat{x}_t)$. Indeed,
\begin{align}
   \hat{J}(t,\bm x):=  & e^{\ii \hat{p}\bm{x}} \delta(\hat{x}_t) e^{-\ii \hat{p}\bm{x}} \nn \\&= \delta ( e^{\ii \hat{p}\bm{x}}\hat{x}_t e^{-\ii \hat{p}\bm{x}})= \delta(\bm{x}-\hat{x}_t).
\end{align}
 The rest of the interaction Hamiltonians that we reviewed in this section can also be brought in this form \eqref{anastop}.

 \subsection{Detector models in curved spacetimes}

The point-like model is straightforwardly adaptable to the case of curved spacetimes. In this case one could treat the interaction to be given by the same Hamiltonian~\eqref{UDWpoint}, but the field amplitude satisfies the Klein-Gordon equation in curved spacetimes and the trajectory is defined over this spacetime background. In this way, the point-like UDW model has been extensively used to analyze aspects of quantum fields in curved spacetimes, such as in cosmology \cite{PhysRevD.15.2738}, and black-hole physics\cite{PhysRevD.89.104002,PhysRevD.90.064003}.  

The definition of extended detectors becomes more subtle in these situations. An extensive account of the impact of spacetime curvature in the formulation of detector models was done recently in \cite{PhysRevD.101.045017, PhysRevD.106.025018}. To define extended detector models in general relativistic set-ups, it is convenient to introduce the interaction Hamiltonian defining the coupling between detector and field in a covariant manner in terms of Hamiltonian densities and Hamiltonian weights~\cite{PhysRevD.101.045017,2020broken}.

In particular in~\cite{PhysRevD.101.045017} an extension of the smeared UDW detector embedded in a general curved spacetime is built. Concretely, the detector is coupled to the field through the following Hamiltonian weight\footnote{The Hamiltonian density $\hat{\mathfrak{h}}$ and the Hamiltonian weight $\hat h$ are related by multiplication by the square root of the determinant of the metric $\hat{\mathfrak{h}}=\sqrt{|g|}\,\hat h$.} in the interaction picture:
\begin{align}\label{hemm}
  \hat h(\mathsf x)=\lambda \Lambda(\mathsf x) \hat {\mu}(\tau(\mathsf x))\otimes\hat \phi(\mathsf x). 
\end{align}
where $\Lambda$ is a space-time function of compact support, $\hat{\mu}$ is again the monopole operator associated with the two-level system, and $\lambda$ is a coupling constant. The idea is that the spacetime smearing function `splits' into the spatial smearing and time switching only in a fixed reference frame, typically chosen to be the Fermi-Walker frame $(\tau,\bm \xi)$ associated with the `centre of mass' of the detector~\cite{PhysRevD.97.105026,PhysRevD.101.045017, PhysRevD.106.025018}. In this frame we could write $\Lambda(\mathsf x)= F(\bm \xi) \chi(\tau)$. Similarly, the detector operator $\hat {\mu}(\tau)$ depends only on the detector's proper time in that frame. Moving to another frame, the detector operator inherits spatial dependence  $ \hat {\mu}(\tau(\mathsf x))$. The covariant notation makes it explicit that the spacetime smearing  $\Lambda(\mathsf x)$ does not specifically belong to the  detector alone or the field alone, and can only be attributed to their joint interaction.

From the Hamiltonian density, one can define a time-dependent Hamiltonian for the joint system as
\begin{align}
    \hat H(\tau)=\int_{\mathcal{E}(\tau)}\!\!\!\!\!\d\mathcal{E} \; \hat h(\mathsf x)
\end{align}
where $\mathcal{E}(\tau)$ is a one-parameter family of space-like surfaces. The parameter $\tau$ is a global function whose level curves represent the planes of simultaneity of the center of mass of the detector, which under some assumptions \cite{PhysRevD.101.045017}, will represent the proper time of the detector.  Finally, $\d\mathcal{E}$ denotes a shortcut for the family of induced measures
\begin{align}
   d\mathcal{E}(\tau) :=\d \mathsf{x}^{n+1}\delta(\tau(\mathsf{x})-\tau)\sqrt{|g|}  (\mathsf{x})
\end{align} in the surfaces $\mathcal{E}(\tau)$.

\subsection{Quantum fields as probes}
Yet another type of detector interaction can be introduced in a totally covariant manner through the introduction of quantum fields as detectors. Indeed, within the literature of detector models we find that one of the two approaches suggested by Unruh in his seminal paper \cite{PhysRevD.14.870} consisted of the coupling of the scalar field to a pair of massive complex scalar fields with different masses, a model that has been used in other contexts \cite{PhysRevD.55.3603}. In the spirit of the previous subsection, these interactions can be expressed in terms of a Hamiltonian weight
\begin{align}\label{huqft}
    \hat h(\mathsf x)=\lambda(\hat\Psi^*_m(\mathsf x)\hat\Psi_M(\mathsf x)+\hat\Psi^*_M(\mathsf x)\hat\Psi_m(\mathsf x))\otimes \hat\phi(\mathsf x).
\end{align}
Field-field couplings were also derived recently through the effective field theory approach in implementations for the experimental verification of the Unruh effect \cite{PhysRevLett.125.213603}.

An extensive study of measurements with local interactions through quantum fields was done recently in the context of algebraic quantum field theory (see~\cite{fewster2020quantum}).
Implicitly, the Hamiltonian weight they considered was given by
\begin{align}\label{hfew}
    \hat h(\mathsf x)=\lambda\Lambda(\mathsf x) \hat\Psi(\mathsf x)\otimes \hat\phi(\mathsf x),
\end{align}
where $\Lambda$ was an arbitrary compactly supported smooth function.

\subsection{Housekeeping}
 Finally, combining ideas from all the models described above, it is possible to unify the description of all these models in a general way, as done in~\cite{PhysRevD.103.085002}. Namely, in what follows, we will be using the following general form of the interaction Hamiltonian weight
\begin{align}\label{h4}
 \hat h(\mathsf x)=\lambda \hat{J}(\mathsf x)\otimes\hat \phi(\mathsf x). 
\end{align}
Here $\hat{J}$ is an arbitrary operator acting over the Hilbert space of the detector. For example, in the pointlike model of the previous section  $\hat{J}(\mathsf x)= \Lambda(\mathsf x) \hat {\mu}(\tau(\mathsf x))$. In  table~\ref{table} below we summarise what the detector current is in every variant that we reviewed in this section.
\begin{table}[ht]
    \centering
    \begin{tabular}{c|c|c}
      Model  & $\hat{J}(\mathsf{x})$ & Microcausality?\\
      \hline
      \eqref{UDWpoint}   & $\displaystyle\int\frac{\d s}{\sqrt{|g(\mathsf{x}(s))|}}\chi(s)\delta(\mathsf{x}-\mathsf x(s))\hat D (s)$&$\checkmark$\\
         \eqref{h1}   & $\chi(t) F(\bm{x})\hat D(t)$&$\times$\\
 \eqref{h3}   & $\chi(t)\delta(\bm{x}-\hat{x}_t)$&$\times$\\
       \eqref{hemm}   & $\Lambda(\mathsf{x})\hat D(\tau(\mathsf{x}))$&$\times$\\
       \eqref{huqft} & $\hat\Psi^*_m(\mathsf x)\hat\Psi_M(\mathsf x)+\hat\Psi^*_M(\mathsf x)\hat\Psi_m(\mathsf x)$&$\checkmark$\\
        \eqref{hfew} & $\Lambda(\mathsf{x})\hat\Psi(\mathsf x)$&$\checkmark$
    \end{tabular}
    \caption{In this table are displayed some of the most popular detector models in terms of their associated current operator $\hat{J}$, and whether they enable superluminal propagation within their respective interaction regions. Models \eqref{h1},\eqref{h2} are defined for inertial reference frames in Minkowski, while the rest can be defined in any reference frame on any curved background since they are explicitly coordinate-free.}
    \label{table}
\end{table}
 Detectors defined in this way may correspond to microcausal as well as nonmicrocausal  detectors. Whether a detector model is microcausal  or not can be characterized by determining whether the dynamics that its interaction Hamiltonian generates propagates information causally. As we studied in \cite{PhysRevD.103.085002}, an equivalent way of characterizing this behavior is to determine whether the associated currents fulfill the microcausality condition, i.e., we say that a detector is non-relativistic if 
 \begin{align}
 [\hat{J}(\mathsf{x}),\hat{J}(\mathsf{y})] \neq 0
\end{align} 
for all spacelike separated  $\mathsf{x},\mathsf{y}$ within the extension of the interaction region. While all the models studied are microcausal in the spatially pointlike limit, when considering finite smearings this is not the case. Assuming some finite spatial smearing, we analyze in Table~\ref{table} which Hamiltonian weights satisfy the microcausality condition.

\section{Quantitative analysis of signalling in general detector models}\label{sec:estimator}
{ In this section we analyze quantitatively the ability of pairs of detectors to signal to each other. This will set the ground to study the notion of approximate spacelike separation by assessing the impact of exponentially suppressed detector smearing tails in signalling in later sections.  Concretely, we will derive, from ideas coming from optimal parameter estimation, a general expression for a signalling estimator between two detectors applicable to all the particle detector variants discussed above.}

{ Let us first set the motivation for this analysis. The concepts of causality and signalling in science are usually discussed in a qualitative way, that is, one does not usually determine (with some explicit quantity)  how much of a cause some event is to some other. To begin this analysis, it should be made clear that causality and signalling are two different notions. Causality involves some notion of agency, e.g. the light in a room went on {\it because} someone turned on the switch. This is a difficult concept to grasp from a statistical physics point of view. In statistics, one is provided with a typically time-dependent probability distribution over a set of parties typically distributed over time, and one can study  the correlations between the observables of these parties.  In order to infer a causal relation,  one needs to be able to parametrize the probability distributions in terms of some controllable variables. Then one can infer the dependence of joint probability distributions (of multipartite systems) on a parameter that is local to only one subsystem. The motto ``correlation does not imply causation'' tells us that the mere observation of  correlations in local observables of both parties does not imply a causal relation between local events associated with each of them. In fact if one does not have control over the value of a local parameter associated with one party in a multipartite system, one cannot infer causal relations between this party and the rest of the system.

Signalling, on the other hand, is understood as a purely statistical concept, related---but not equivalent---to causality. One associates with each party a set of possible measurement outcomes and a set of `settings' or parameters (as it is done e.g. in the process matrix formalism \cite{PhysRevX.8.011047}). It is said that a party A can signal another party B if the local statistics of B depends on the local settings of A. The relevant statistics for signalling is the local statistics associated with each party. In particular, two parties can be correlated and yet not be able to interchange signals. On the other hand, it could be that there exist causal relations that do not translate into signalling in the statistics, in the presence of fine-tuning and feedback loops~\cite{https://doi.org/10.48550/arxiv.2203.11245}. These scenarios where causal relationships appear in the absence of signalling are rather contrived. For the case of particle detecctor models interacting in the ways specified in Section~\ref{sec:zoo}, the notion of signalling is tied to the notion of relativistic causality and as such, within this framework (without modifying the model) the presence of signalling from parameters of detector A to local observables of detector B is sufficient (and very likely necessary) for detector A to causally influence detector B. In this work, we will focus primarily on signalling relations. 

One can wonder how the statistical notion of causality relates with causality as defined in relativistic theories. In relativity, one represents spacetime by a smooth manifold with a Lorentzian metric. In this manifold one defines a partial order relation: It is said that a point in a manifold precedes another if they can be connected through a future-directed causal (non-spacelike) curve.  For a given point $p$, the set of all such points is called the causal future of $p$, denoted $\mathcal{J}^+(p)$. 
 Relativistic causality generally involves a set of parties embedded in the spacetime manifold. The causal relations between parties should be compatible with the partial order implied by the causal structure of the manifold, in such a way that one party cannot causally influence another if they are in spacelike separated regions. It is common that, in practice, a less restrictive (but physically relevant) constraint is imposed between the parties. Namely, that two parties in spacelike separation cannot send signals to each other. 
 
 In the case of relativistic QFT, the compatibility of the spacetime causal structure with the (statistical) signalling relations between parties---which are associated with compact regions of spacetime---is a consequence of the microcausality condition. For example, for a scalar field $\hat\phi$ this is 
 \begin{align}
     [\hat\phi(\mathsf{x}),\hat\phi(\mathsf{y})]=0
 \end{align}
when $\mathsf{x}$ and $\mathsf{y}$ are spacelike separated points. Roughly speaking, since all observables and operations are built from the field operators, the microcausality condition would imply the following: if $\hat\rho_{O}$ is a local state built from operators localizable in a region $O$ and $\Phi_{O'}$ is a CPTP map built from operators localizable in a region $O'$ spacelike separated from region $O$ then
\begin{align}
    \Phi_{O'}[\hat\rho_{O}]=\hat\rho_{O}.
\end{align}
Therefore, the local operation $\Phi_{O'}$ cannot affect the local statistics that are encoded in $\hat\rho_{O}$. In this sense, the microcausality condition can be read as a non-signalling condition. 

The simple link between signalling and microcausality in QFT we just exposed becomes more subtle  when tackling QFT calculations more rigorously~\cite{earman2014relativistic}. Indeed, relativistic constraints preclude the existence of local states such as $\hat\rho_{O}$. Restrictions of global states, e.g. the vacuum state, to local states typically result in states that are not trace-class, and therefore the picture of signalling in terms of maps acting over local states breaks. These subtleties can be addressed, for example, by restricting the ability of observers to act on the field locally to the coupling of local probes modelled by particle detectors~\cite{PhysRevD.105.065003}.

\subsection{Signaling between two particle detectors coupled to the field}

To characterize signalling between two particle detectors, and discuss its compatibility with relativistic causality, we first consider the time evolution operator $\hat U_{\textsc{a}+\textsc{b}}$ generated by the interaction Hamiltonian of two detectors A and B and the field. We say that  detector A can signal detector B if the reduced state of B after the interaction,
\begin{align}
    \hat\rho_\textsc{b}=\tr_{\textsc{a},\phi}(\hat U_{\textsc{a}+\textsc{b}}\;\hat\rho_{\text{initial}}\;{\hat U}^{\dagger}_{\textsc{a}+\textsc{b}}),
\end{align}
(where $\tr_{\textsc{a},\phi}$ denotes trace over the degrees of freedom of detector A and the field) depends on any parameter of detector A. One can say that the model is compatible with relativistic causality if the signalling allowed by the model is compatible with the relativistic causal relations associated with the respective interaction regions. 

So far, these are all qualitative notions. Based on the definition of signalling given above, we can determine whether a detector can or cannot signal another detector, but we have not introduced a way of determining `how much'. Such analysis is important on its own (to assess the causal behaviour of detector models as a whole), but it becomes crucial in situations where interactions are not strictly localized in spacetime (which as discussed in Section~\ref{sec:zoo} are also physically relevant), and one therefore has no way of imposing relativistic causality because the detectors are not associated with compact regions. The quantitative analysis of signalling between non-compact detectors was first discussed in~\cite{PhysRevD.92.104019}.  Let us outline the ideas and results in that reference. In~\cite{PhysRevD.92.104019} the analysis was restricted to the case of interaction Hamiltonians of the form~\eqref{h1}, that is, smeared UDW detectors. More concretely, it was considered that two detectors are coupled to the field through the following interaction Hamiltonian 
\begin{align}
    \hat{H}= \sum_{\nu}\lambda_{\nu} \chi_{\nu}(t)\hat{\mu}_{\nu}(t)\otimes \int \d \bm{x}\, F_{\nu}(\bm{x})\hat{\phi}(t,\bm x)
\end{align}
where $\nu=$A, B. The overall picture is the following: the two detectors interact with the quantum field, initially being in an uncorrelated state. After the interaction, the detectors will be correlated, and the goal is to distil the influence that one of the detectors (the sender) has over the local statistics of the other (the receiver) at leading order in perturbation theory. In \cite{PhysRevD.92.104019} there were some extra assumptions made.  The switching functions $\chi_\textsc{a}(t),\chi_\textsc{b}(t)$ were considered to be compactly supported over disjoint time intervals, so that one detector was set to interact with the field strictly after the other in the particular reference frame in which the global interaction was defined. The smearing functions, however, were not necessarily compactly supported. 

Calculating the effect of the interaction at leading order in perturbation theory, the perturbed dynamics of the density operator of the receiver B could be split into two terms: one of them carrying the parameters associated with the sender A (dubbed the signalling part) and another independent of these (dubbed the noise part). More concretely, the leading order correction to the density matrix of the receiver was found to be
\begin{align}
    \hat\rho^{(2)}_{\textsc{b}}= \lambda_{\textsc{a}} \lambda_{\textsc{b}} \hat{\rho}^{(2)}_{\textsc{b},\text{signal}}+ \lambda^2_{\textsc{b}} \hat{\rho}^{(2)}_{\textsc{b},\text{noise}}, \label{rho2}
\end{align}
where the noise term is local on detector B, and all the influence of the presence of detector A on detector B's density matrix is captured by the term
\begin{align}
  & \nn \hat\rho^{(2)}_{\textsc{b},\text{signal}}\\
    &= \int \d t \int \d t' \chi_\textsc{a}(t)\chi_\textsc{b}(t')  \text{Tr} [\hat{\mu}_\textsc{a}(t) \hat\rho_\textsc{a}]\mathcal {C}(t,t')[\hat{\mu}_\textsc{b}(t'),\hat\rho_\textsc{b}] \label{edu}
\end{align}
with 
\begin{align}
    \mathcal{C}(t,t')=\ii \int \d \bm{x} \int \d \bm{x'} F_\textsc{a}(\bm{x})F_\textsc{b}(\bm{x'}) \langle [\hat{\phi}(t,\bm x), \hat{\phi}(t',\bm x')] \rangle. \label{com}
\end{align}

First, note that if the two interaction regions are not causally connected, the smeared field commutator vanishes due to the microcausality condition, and thus so does $\hat\rho^{(2)}_{\textsc{b},\text{signal}}$. In other words, microcausality guarantees no FTL signalling at leading order\footnote{Moreover, it can be shown non-perturbatively that if the interactions are causally disjoint there is no faster than light signalling between pairs of detectors \cite{PhysRevD.103.085002}}. {Of course, if the interaction regions are not bounded in spacetime, the two detectors can be in causal contact even if the centre of the interaction regions are spacelike separated. This does not imply faster-than-light signalling, since the signal exchanges between the detectors will come from the causal interactions between the respective tails. We will denote the information carried by these signals as {``apparent''} FTL signalling. For example, if two detectors are Gaussian localized and their centers of mass are spacelike separated by a proper length of more than, say, seven times their standard deviation, it is often considered in the literature that the detectors are effectively spacelike separated. Here we will analyze quantitatively how accurate that kind of statements are.

Second, note that the state of the field does not affect the signalling term in the perturbed dynamics (for free quantum fields). This is because the distribution $\mathcal{C}(t,t')$ is independent of the field's state, and therefore, at leading order, the state of the field does not play a relevant role in signalling.

In what follows we are going to generalize the  results in~\cite{PhysRevD.92.104019} for the general detector model~\eqref{h4}. We will also introduce a novel definition of a signalling estimator based on quantum informational notions and optimal parameter estimation. This will allow us to discriminate whether the effects of exponential tails are relevant enough to break the ``apparent'' localization of the detectors in spacetime on a case-by-case basis. }}

\subsection{Joint dynamics of pairs of detectors}

In order to analyze signalling relations between pairs of detectors in the most general set-up described in section \ref{sec:zoo}, we will calculate the general leading order behavior of the state of detector B (again, the receiver) interacting weakly with a quantum field that has previously interacted with another detector A (the sender).

Consider the case of two general detectors, A and B, which interact with the field according to the interaction Hamiltonian
\begin{align}
   \sum_{\nu=\text{A,B}} \hat H_{\nu}(\tau)=\sum_{\nu=\text{A,B}}\int_{\mathcal{E}(\tau)}\!\!\!\!\!\d\mathcal{E}\;   \hat h_{\nu}(\mathsf x).
\end{align}
where in this case the corresponding Hamiltonian weights will be given by\footnote{Recall that, for convenience, we have absorbed the spacetime smearing function in the definition of the detector current operator, e.g., $\hat{J}_{\nu}(\mathsf x):=\Lambda_{\nu}(\mathsf x)\hat{\mu}_{\nu}(\mathsf x (\tau))$.}
\begin{align}
    &\hat h_\textsc{a}(\mathsf x)=\lambda_\textsc{a} \hat J_\textsc{a}(\mathsf x)\otimes\openone_\textsc{b}\otimes\hat \phi(\mathsf x) \label{haJ}
\end{align}
and
\begin{align}
    &\hat h_\textsc{b}(\mathsf x)=\lambda_\textsc{b}  \openone_\textsc{a}\otimes\hat J_\textsc{b}(\mathsf x)\otimes\hat \phi(\mathsf x). \label{hbJ}
\end{align}
  The joint evolution in the interaction picture of the detectors and the field can be described as a unitary operator acting over the joint initial state of the field-detectors system $\hat\rho_{\text{initial}}$. Then the state in the asymptotic future will be given by the transformation
\begin{align}
    \hat\rho_{\text{final}}=\hat U_\textsc{a+b}\;\hat\rho_{\text{initial}}\;{\hat U}^{\dagger}_\textsc{a+b}, \label{19}
\end{align}
where $\hat U_\textsc{a+b}$ is the evolution generated by the detector-field interaction Hamiltonian (see Appendix~\ref{app:sigma}). The local statistics of the detector B will be given by the partial trace
\begin{align}
    \hat\rho_\textsc{b}=\tr_{\textsc{a},\phi}(\hat U_\textsc{a+b}\;\hat\rho_{\text{initial}}\;{\hat U}^{\dagger}_\textsc{a+b}).
\end{align}
and the signalling term can be defined as 
\begin{align}
  & \nn \hat\rho^{(2)}_{\textsc{b},\text{signal}}\\
  &=  \frac{\partial^2}{\partial{\lambda_\textsc{a}}\partial{\lambda_\textsc{b}}}\tr_{\textsc{a},\phi}(\hat U_{\textsc{a+b}}\;\hat\rho_{\text{initial}}\;{\hat U}^{\dagger}_{\textsc{a+b}})|_{\lambda_\textsc{a}=\lambda_\textsc{b}=0} . \label{estimator}
  \end{align}
similarly to equation \eqref{rho2} above. Using the Dyson expansion, in appendix \ref{app:sigma} we derive that 
\begin{align}
    \hat\rho^{(2)}_{\textsc{b},\text{sign}}=-\ii[\hat{\Sigma},\hat\rho_\textsc{b}] \label{sigma1!}
\end{align}
where we have defined the operator
\begin{align}
    \nn&\hat{\Sigma}=\int\int\d V\d V' \braket{\hat J_\textsc{a}( \mathsf x')}G_\textsc{r}(  \mathsf x,\mathsf x')\hat J_\textsc{b}( \mathsf x)\\
    &=\int\d V G_{\textsc{r}}[\braket{\hat J_\textsc{a}}](  \mathsf x)\hat J_\textsc{b}( \mathsf x). \label{sigma2!}
\end{align}
Here $G_\textsc{r}(\mathsf x,\mathsf x')$ is the retarded Green function 
\begin{align}
   G_\textsc{r}(  \mathsf x,\mathsf x')=-\ii \theta(\tau(\mathsf{x})-\tau(\mathsf{x'}))\braket{[\hat\phi(\mathsf{x}),\hat\phi(\mathsf{x}')]}\label{sigma3!}.
\end{align}
$\d V$ denotes the invariant element of volume with respect to the background metric
\begin{align}
    \d V=\d \mathsf{x}^{n+1}\sqrt{|g|},
\end{align}
where $|g|$ is the (absolute value of the) determinant of the metric. 

The operator $\hat{\Sigma}$ can be understood as the current associated with detector B smeared by the propagated expectation value of the current associated with detector A, which we defined as
\begin{align}
   G_{\textsc{r}}[\braket{\hat J_\textsc{a}}](  \mathsf x):= \int\d V' \braket{\hat J_\textsc{a}( \mathsf x')}G_\textsc{r}(\mathsf x,\mathsf x'). \label{sigma4!}
\end{align}
In the case of the massless Klein-Gordon field in a 3+1 dimensional flat spacetime, for instance, the propagator takes the familiar form of the Lienard-Wiechert potentials
\begin{align}\label{lienardwiechert}
    G_{\textsc{r}}(\mathsf x,\mathsf x')=\theta(x_0-x'_0)\delta\left[(\mathsf x-\mathsf x')^2\right]
\end{align}
and
\begin{align}
     G_{\textsc{r}}[\braket{\hat J_\textsc{a}}](t,   \bm x)=\int \d^3\bm x' \frac{\braket{\hat J_\textsc{a}}(t_{\textsc{r}},\bm {x}' )}{2|\bm {x}-\bm {x}'|}
\end{align}
where $t_{\textsc{r}}=t-|\bm {x}'|$ is the retarded time.

The above results \eqref{sigma1!}-\eqref{sigma4!} generalize \eqref{edu}. In \cite{PhysRevD.92.104019} it was assumed that the switching functions were compactly supported and non-overlapping, so one interaction happens `after' the other in some reference frame. By dropping this assumption, for general detector models, the role of the field commutator is played by the field's retarded Green's function \eqref{sigma3!}.

\subsection{Quantum Fisher information and a generalized signalling estimator}

At this point we are ready to generalize the signalling estimator defined in \cite{PhysRevD.92.104019} for general detector models, beyond the assumption of compact support for the interaction and for general globally hyperbolic spacetimes.  The definition of the operator $\hat{\Sigma}$ in \eqref{sigma2!}, without further assumptions, should be regarded as merely formal. Indeed, note that, given an arbitrary globally hyperbolic space-time, the retarded Green's function $G_\textsc{r}$ is guaranteed to exist as an ordinary distribution acting over compactly supported functions \cite{frankwave}.  Therefore, it is not a priory guaranteed that the expression \eqref{sigma2!} makes sense if, e.g., the mean value of $\hat J_\textsc{a}( \mathsf x')$ is not compactly supported. However, it is known that this is not a problem in many of the common cases studied in the literature whenever there are no infrared ambiguities in the theory. 

In the spirit of~\cite{PhysRevD.92.104019} one could be tempted to define a signalling estimator as the norm of the operator $\hat{\Sigma}$ in~\eqref{sigma2!}. While this would be always well defined for finite dimensional particle detectors, this may not be well defined for more general models. In particular the operator $\hat{\Sigma}$ involves the smearing of the operator $\hat J_\textsc{b}( \mathsf x)$, which does not have well-defined support as an operator in general. To build a meaningful signalling estimator for the general case, we are forced therefore to specify the particular configuration of the state of the detectors.

To build a signalling estimator we will analyze the issue of signalling in detector models from the perspective of quantum metrology~\cite{Paris}, which is precisely concerned with the estimation of a parameter that is (dynamically) encoded in the state of a quantum system. 
Namely, we will claim that there is no signalling if B cannot access the parametric information that is encoded in the state of detector A after its interaction with the field. To make it more concrete, we will establish that there will be no signalling if B cannot infer the value of the coupling constant $\lambda_\textsc{a}$ through its local statistics.

A core concept in parameter estimation in quantum metrology is the so-called quantum Fisher information~\cite{Paris,Luo2000}. Given a family of density matrices that are dependent on some parameter, say $\lambda$, the quantum Fisher information yields lower bounds on the variance of the distribution of possible values of $\lambda$ given some certain measurement statistics on the system. When the quantum Fisher information is close to zero the statistical variance of the optimal parameter estimation grows to infinity, which is a consequence of the Cramer-Rao bound~\cite{Paris}. 
In our case, we are going to consider the dependence of the partial state of detector B on $\lambda_\textsc{a}$ so that the quantum Fisher information estimates how much the information about whether A coupled to the field or not (and how strongly) is accessible to B. Note that the whole influence of any parameter of A on B is conditional to the coupling constant $\lambda_\textsc{a}$ being nonzero.

Given a one-parametric family of density matrices \mbox{$\hat{\rho}(\lambda)=\sum p_i(\lambda) \ket{i(\lambda)}\!\!\bra{i(\lambda)}$}, the quantum Fisher information at $\lambda$ is given by \cite{Paris}:
\begin{align}\label{fisher}
    \mathcal{F}(\lambda)=2\!\!\!\sum_{\{p_l+p_m>0\}}\frac{\left|\braket{l|\partial_\lambda \hat \rho(\lambda)|m}\right|^2}{p_l+p_m}.
\end{align}
Since our parameter will be the coupling strength of detector A and our signalling estimation will be defined around the regime of weak couplings, we can expand the Fisher information at the lowest orders in the coupling constants and compute the Fisher information around the value $\lambda_\textsc{a}=0$.
To be more explicit, we are interested in the estimation of the parameter $\lambda_\textsc{a}$ (close to zero) from the local statistics of B after the interaction,  at leading order in perturbation theory. From equations \eqref{rho2}, \eqref{sigma1!}, we find that 
\begin{align}
    \partial_{\lambda_\textsc{a}}\hat\rho_\textsc{b}(\lambda_\textsc{a})|_{\lambda_\textsc{a}=0}=-\ii\lambda_\textsc{b}[\hat{\Sigma},\hat\rho_\textsc{b}]+\mathcal{O}(\lambda_\textsc{b}^2), \label{partialA}
\end{align} 
  It is easy to see by direct substitution of \eqref{partialA} in \eqref{fisher} that the expression for the Fisher information of detector B with respect to the parameter $\lambda_{\textsc{a}}$ is given by
\begin{align}\label{fisher2}
    \mathcal{F}_\textsc{b}(\lambda_\textsc{a})|_{\lambda_\textsc{a}=0}=2\lambda_\textsc{b}^2\!\!\!\!\!\!\!\sum_{\{p_l+p_m>0\}}\frac{(p_l-p_m)^2}{p_l+p_m}\left|\braket{l|\hat{\Sigma}|m}\right|^2+\mathcal{O}(\lambda_\textsc{b}^3).
\end{align}
Note that at leading order, the change on $\hat\rho_\textsc{b}(\lambda_{\textsc{a}})$ with $\lambda_{\textsc{a}}$ is given by taking the commutator with the self-adjoint operator $\hat{\Sigma}$, which means that at leading order the dependence on $\lambda_{\textsc{a}}$ is given by the action of a unitary with generator $\hat{\Sigma}$.  When the family of density matrices is generated by the action of a unitary group over a pure state, it holds that the quantum Fisher information coincides with four times the variance of the generator of the unitary ~\cite{Luo2000}. This means that, at leading order,
\begin{align}\label{fishervar}
&    \mathcal{F}_\textsc{b}(0)= 4\lambda_\textsc{b}^2\left(\braket{\psi_\textsc{b}| \hat{\Sigma}^2|\psi_\textsc{b}}-\braket{\psi_\textsc{b}| \hat{\Sigma}|\psi_\textsc{b}}^2\right)+\mathcal{O}(\lambda_\textsc{b}^3).
\end{align}
When the unitary group acts over general mixed states, the variance gives an upper bound for the quantum Fisher information~\cite{Luo2000}
\begin{align}\label{fishervar2}
&    \mathcal{F}_\textsc{b}(0)\leq 4\lambda_\textsc{b}^2\left(\braket{ \hat{\Sigma}^2}_{\hat\rho_\textsc{b}}-\braket{ \hat{\Sigma}}_{\hat\rho_\textsc{b}}^2\right)+\mathcal{O}(\lambda_\textsc{b}^3).
\end{align}

Therefore, we can define the following signalling estimator 
\begin{align}
    \mathcal{S}=\braket{ \hat{\Sigma}^2}_{\hat\rho_\textsc{b}}-\braket{ \hat{\Sigma}}_{\hat\rho_\textsc{b}}^2, \label{y}
\end{align}
 generalizing the estimator defined in \cite{PhysRevD.92.104019} to unbounded operators. in our case, the Fisher information estimates signalling by bounding the information that one detector can ``learn'' about the coupling of other detectors to the same quantum field.
Note that \eqref{y} is exact (at leading order) for initially pure detector states and  provides an upper bound if the initial states are non-pure, as in \eqref{fishervar2}. 

For the purposes of quantifying whether the setup is devoid of apparent superluminal signalling, having an upper bound to the Fisher information is enough. Concretely, the predictions of a given model are reliable if the upper bound is sufficiently `small' given the initial states of the detectors and the choice of background spacetime. In this sense, this estimator defines the regime of validity of each model. Nevertheless, this estimator may not be faithfully estimating the amount of signalling that the sender can transmit to the receiver, if simultaneously the variance is large, the detectors' operators are not bounded and the initial states are mixed. In that case, to obtain a faithful measure of signalling one would have to calculate the actual Fisher information (i.e., not only its upper bound) which can be involved depending on the model under consideration. 

Substituting Eq.~\eqref{sigma2!} into the signalling estimator~\eqref{y}, we get 
\begin{align}\label{Siggy}
    &\mathcal{S}=\\
    \nn &\frac{1}{2}\iint \d V\d V'\; G_{\textsc{r}}[\braket{\hat J_\textsc{a}}](  \mathsf x)  G_{\textsc{r}}[\braket{\hat J_\textsc{a}}](\mathsf x')  \braket{\{\hat J_\textsc{b}(\mathsf{x}),\hat J_\textsc{b}(\mathsf{x'})\}}\\
\nn & -(G_\textsc{r}[ \braket{\hat J_\textsc{b}},\braket{\hat J_\textsc{a}}])^2, 
\end{align}
where $\{\cdot,\cdot\}$ denotes the anticommutator and  \begin{equation}
  G_\textsc{r}[ \braket{\hat J_\textsc{b}},\braket{\hat J_\textsc{a}}]  = 
  \int\d V G_{\textsc{r}}[\braket{\hat J_\textsc{a}}](  \mathsf x)\langle \hat J_\textsc{b}( \mathsf x) \rangle \label{s}
\end{equation}
 is the `overlap' of the expectation values of the currents, convolved with the retarded Green's function.

This estimator captures the main features outlined in \cite{PhysRevD.92.104019} to quantify signalling.  A first consistency check is that, indeed, if the functions  $\braket{\hat J_\textsc{a}}(\mathsf{x})$, $\braket{\hat J_\textsc{b}}(\mathsf{x})$, and $\braket{\{\hat J_\textsc{b}(\mathsf{x}),\hat J_\textsc{b}(\mathsf{x'})\}}$ are compactly supported, and if the support of these functions are space-like separated with respect to each other, then the estimator is zero, i.e., there is no signalling between strictly spacelike separated detectors. 

The signalling estimator will not be zero if  these functions are not compactly supported, but one would expect that detectors that are, in some sense, ``centered'' around a region cannot significantly influence events outside the future lightcone of this region. Therefore, for any notion of effective localization  of a detector, we can define a notion of effective lightcone based on the signalling estimator \eqref{Siggy}.  Roughly speaking, two detector interactions ``centered'' in spacelike separation can be considered to be effectively spacelike separated if the estimator is negligible. 

To make sense of this definition, a more elaborate definition of an interaction ``centered'' at a point in spacetime is required. The estimator \eqref{Siggy} involves the mean values and the fluctuations of the current operators $\hat J_{\textsc{a},\textsc{b}}$, which are the only values that can affect the detectors' statistics at quadratic order in perturbations. Therefore, these are the relevant functions that determine the localization of the detectors at that order.  We can define these functions to be centered in a region if, e.g., these functions decay rapidly away from any point belonging to this region. We shall discuss these effective notions of localization in next section.

\section{Expression of the signalling estimator in some particular cases}\label{sec:examples}
In this section we provide expressions for our signalling estimator \eqref{y} in some cases of interest, to gain some intuition about the physics that it captures in each case. In principle, for every detector model one must consider the dependence of the signaling estimator on all parameters of the interactions, the initial states of the detectors, and the different dynamics of the field (for each possibly curved spacetime background). Such an exhaustive analysis for all types of detector models in curved spacetimes, and the impact that effective localization has on causality in each case, is outside the scope of this work. We will simply comment on some aspects of the signaling estimator in the examples below. In the next section, we will perform a more in-depth analysis of the simplest case (signaling between two-level systems).

\subsection{Signalling between two two-level systems}
Consider the case of two Unruh-DeWitt detectors that interact with a quantum field in a curved spacetime background. In the case of Unruh-DeWitt detectors, the currents can always be written covariantly as
\begin{align}
     \hat J_{\nu}=\sum_{s={\pm 1}}\Lambda^{s}_\nu\hat{\sigma}^s_\nu
\end{align}
where the detector index $\nu\in\{\text{A},\text{B}\}$ and where we have defined
\begin{align}\label{lambdaboost}
    \Lambda^{s}_\nu(\tau_\nu,\bm {x}_\nu)\coloneqq\chi_\mu(\tau_\nu)F_\nu(\bm {x}_\nu)e^{\ii s\Omega_\nu \tau_\nu}
\end{align}
for two different set of coordinates $(\tau_\nu,\bm {x}_\nu)$ for $\nu\in\{\text{A,B}\}$. The operators  $\hat{\sigma}^{\pm1}_{\nu}\equiv\hat{\sigma}^{\pm}_{\nu}$ represent the ladder operators associated with each two-level system.

Given this decomposition, the operator $\hat\Sigma$ can be written as
\begin{align}
    &\nn\hat\Sigma=\sum_{s,s'={\pm 1}}\hat{\sigma}^s_\textsc{b}\braket{\hat{\sigma}^s_\textsc{a}}G_{\textsc{r}}[\Lambda^{s}_\textsc{b}, \Lambda^{s'}_\textsc{a}]\\
    &=\sum_{s,s'={\pm 1}}\hat{\sigma}^s_\textsc{b}\braket{\hat{\sigma}^s_\textsc{a}}I(s'\Omega_{\textsc{a}},s\Omega_{\textsc{b}}), \label{sigmaudw}
\end{align}
where we have defined 
\begin{align}\label{Iboost}
I(s'\Omega_{\textsc{a}},s\Omega_{\textsc{b}}):=G_{\textsc{r}}[\Lambda^{s}_\textsc{b}, \Lambda^{s'}_\textsc{a}].
\end{align}
Then, from the definition \eqref{y}, the signaling estimator for this model is given by the variance of the operator $\hat{\Sigma}$ in equation \eqref{sigmaudw}
\begin{align}\label{signUDWtotalfreq}
\nn\mathcal{S} =   \sum_{s,s'=\pm 1}&\langle\hat{\sigma}^s_{\textsc{a}}\rangle\langle\hat{\sigma}^{s'}_{\textsc{a}}\rangle \bigg[I(s\Omega_{\textsc{a}},\Omega_{\textsc{b}})I(s'\Omega_{\textsc{a}},-\Omega_{\textsc{b}})\\
& -\sum_{u,u'=\pm 1}\langle\hat{\sigma}^u_{\textsc{b}}\rangle\langle\hat{\sigma}^{u'}_{\textsc{b}}\rangle I(s\Omega_{\textsc{a}},u\Omega_{\textsc{b}})I(s'\Omega_{\textsc{a}},u\Omega_{\textsc{b}})\bigg].
\end{align}


Note that, in the case of two-level systems, one can maximize the signalling between the detectors for all possible states (the optimal value $\mathcal{S}_{max}$ will play a fundamental role in later sections). Indeed, we notice that the second term is negative and vanishes for $\langle\hat{\sigma}^{\pm}_{\textsc{b}}\rangle=0$. Then, if we write the expectation value of the ladder operators associated with detector A in polar form, i.e. $\langle\hat{\sigma}^{\pm}_{\textsc{a}}\rangle=re^{\pm\ii \alpha}$ with $0\leq r\leq1$, we realize that the first term can be written as
\begin{align}
\nn\mathcal{S}&=r^2\left|e^{\ii\alpha}I(\Omega_{\textsc{a}},-\Omega_{\textsc{b}})+e^{-\ii\alpha}I(\Omega_{\textsc{a}},\Omega_{\textsc{b}})\right|^2\\
    \nn& =r^2\left(|I(\Omega_{\textsc{a}},-\Omega_{\textsc{b}})|^2+|I(\Omega_{\textsc{a}},\Omega_{\textsc{b}})|^2\right.\\
    &\left.+2\Re \; e^{\ii2\alpha}I(\Omega_{\textsc{a}},\Omega_{\textsc{b}})I^*(\Omega_{\textsc{a}},-\Omega_{\textsc{b}})\right). \label{adependence}
\end{align}
We see that in the signalling estimator there is a contribution coming from an interference term. If we write this interference term in polar form, that is
\begin{align}\label{inconsequentialphase}
    I(\Omega_{\textsc{a}},\Omega_{\textsc{b}})I^*(\Omega_{\textsc{a}},-\Omega_{\textsc{b}})=e^{-\ii\beta}|   I(\Omega_{\textsc{a}},\Omega_{\textsc{b}})||I(\Omega_{\textsc{a}},-\Omega_{\textsc{b}})|, 
\end{align}
where $\beta$ is just the principal argument of the complex number \eqref{inconsequentialphase}. The signalling estimator can be written then as
\begin{align}
    \nn\mathcal{S}&=r^2\left(|I(\Omega_{\textsc{a}},-\Omega_{\textsc{b}})|^2+|I(\Omega_{\textsc{a}},\Omega_{\textsc{b}})|^2\right.\\
   &\left.+2 \cos (\alpha-\beta)|   I(\Omega_{\textsc{a}},\Omega_{\textsc{b}})||I(\Omega_{\textsc{a}},-\Omega_{\textsc{b}})|\right), 
\end{align}
which is maximum when setting $r=1$ and $\alpha=\beta$.

Therefore, the maximum value for the signalling estimator is achieved when the state of the detector B is diagonal in the basis of its free Hamiltonian, and the state of detector A is such that $\langle\hat{\sigma}^{\pm}_{\textsc{a}}\rangle=e^{\ii \beta}$. For these states,  the maximum value of the signalling estimator takes the form 
\begin{align}\label{MaxSignalling}
\mathcal{S}_{\text{max}}=\left(|I(\Omega_{\textsc{a}},-\Omega_{\textsc{b}})|+ |I(\Omega_{\textsc{a}},\Omega_{\textsc{b}})|\right)^2.
\end{align}

\subsection{Signaling between two quantum particles}
One can also consider models of the Unruh-Wald type, which describes the interaction between two spinless charged particles through a scalar field and is modeled by Hamiltonians like
\begin{align}
\hat{H}_{\textsc{uw}}= \sum_{\nu\in\{A,B\}} \lambda_{\nu} \chi_{\nu}(t) \int \d^n \bm{x} \hat{\phi}(t,\bm x) \delta (\bm{x}-\hat{\bm x}_{\nu}(t)).
\end{align}
In that case, 
\begin{align}
    \hat{J}_{\nu}(\mathsf{x})=\chi_{\nu}(t)\delta (\bm{x}-\hat{\bm x}_{\nu}(t)),
\end{align}
which leads to
\begin{align}
   &\nn\braket{ \hat{J}_{\textsc{a}}}(\mathsf{x})=\chi_{\textsc{a}}(t)\braket{\delta (\bm{x}-\hat{\bm x}_{\textsc{a}}(t))}\\
   &=\chi_{\textsc{a}}(t)|\psi_{\textsc{a}}(t,\bm x)|^2, \label{Ja}
\end{align}
where $\psi_{\textsc{a}}(t,\bm x)$ is the A's wave function, and
\begin{align}
    &\nn\braket{\hat J_\textsc{b}(\mathsf{x})\hat J_\textsc{b}(\mathsf{x'})}\\
    &\nn=\chi_\textsc{b}(t)\chi_\textsc{b}(t')\braket{\delta (\bm{x}-\hat{\bm x}_\textsc{b}(t))\delta (\bm{x}'-\hat{\bm x}_\textsc{b}(t'))}\\
    &=\chi_\textsc{b}(t)\chi_\textsc{b}(t')\psi_\textsc{b}^*(t,\bm x)\psi_\textsc{b}(t',\bm x')\mathcal{G}_\textsc{b}(t-t', \bm x,\bm x'),
\end{align}
where 
\begin{align}
    \mathcal{G}_\textsc{b}(t-t',\bm x,\bm x')=\braket{\bm x|\hat U_{\textsc{b},\text{free}}(t-t')|\bm x'}.
\end{align}
$\hat U_{\textsc{b},\text{free}}$ is the unitary operator associated with the internal (uncoupled) dynamical evolution of B, and therefore $\mathcal{G}_\textsc{b}$ is the uncoupled propagator of detector B.  In contrast to the previous case, unless there are further constraints, there is no optimization with respect to the states that will set the second term of \eqref{Siggy} to zero, since 
\begin{align}
     G_\textsc{r}[ \braket{\hat J_\textsc{b}},\braket{\hat J_\textsc{a}}]  &= \int \d t \d \bm{x} \chi_{\textsc{b}}(t)|\psi_{\textsc{b}}
     (t,\bm{x})|^2\nn \\ &\times \int \d t' \d \bm{x}' G_\textsc{r} (t,\bm{x},t',\bm{x}') \chi_{\textsc{a}}(t') |\psi_{\textsc{a}}(t',\bm{x}')|^2
\end{align}
is non-zero in general for valid normalized states of A, B. We can still derive a bound from the first term of \eqref{Siggy} we derive the bound
\begin{align}
    &\mathcal{S} \leq \int  \d t \d \bm{x}  \int \d t' \d \bm{x}'G_\textsc{r}[\langle \hat{J}_{\textsc{a}}\rangle ](t,\bm x) G_\textsc{r}[\langle \hat{J}_{\textsc{a}}\rangle ](t',\bm x') \nn \\
    &\times \mathcal{G}_{\textsc{b}}(t-t',\bm{x},\bm{x}')\chi_\textsc{b}(t) \chi_\textsc{b}(t') \psi_\textsc{b}^*(t,\bm x) \psi_\textsc{b}(t',\bm x') \label{ti}
\end{align}
where 
\begin{align}
    G_\textsc{r}[\langle \hat{J}_{\textsc{a}}\rangle ](t,\bm x)= \int \d t' \d \bm{x}' G_{\textsc{r}}(t,\bm{x},t', \bm{x}')\chi_\textsc{a}(t') |\psi_\textsc{a}(t',\bm x')|^2. \label{forw}
\end{align}

We see that in this case $ G_\textsc{r}[\langle \hat{J}_{\textsc{a}}\rangle ]$ is a convolution of the field's retarded propagator, detector A's switching function and the probability density of A. If A's wavefunction at $t=0$ is compactly supported on a spatial region $\Delta$ and detector A interacts with the field only at $t=0$ with a delta switching function $\chi_\textsc{a}(t)= \delta (t)$ then 
\begin{equation}
     G_\textsc{r}[\langle \hat{J}_{\textsc{a}}\rangle ](t,\bm x)= \int_{\Delta} \d \bm{x}' G_{\textsc{r}}(t,\bm{x},0, \bm{x}')|\psi_\textsc{a}(0,\bm x')|^2
\end{equation}
is compactly supported in the causal future of $\Delta$. On the other hand, if the switching function has a non-zero extension in time e.g. a finite time extension $\epsilon>0$, $G_\textsc{r}[\langle \hat{J}_{\textsc{a}}\rangle ](t,\bm x)$ is supported on the whole $t>0$ plane due to the instantaneous spreading of A's wavefunction \cite{Hegerfeldt:1998ar}, i.e., the fact that $\psi_\textsc{a}(\epsilon,\bm x')$ has support everywhere for all $\epsilon>0$. The upper bound of \eqref{ti} can be written as
\begin{align}
   \mathcal{S} \leq \int  \d t \d \bm{x}  \int \d t' \d \bm{x}' \psi_\textsc{ab} (t',\bm x')  \mathcal{G}_{\textsc{b}}(t-t',\bm{x},\bm{x}') \psi^*_\textsc{ab} (t,\bm x) 
\end{align}
where 
\begin{equation}
    \psi_\textsc{ab} (t,\bm x):=  G_\textsc{r}[\langle \hat{J}_{\textsc{a}}\rangle ](t,\bm x) \chi_\textsc{b}(t) \psi_\textsc{b}(t,\bm x) 
\end{equation}
This is the `overlap' of the propagated current of A \eqref{Ja} with B's wavefunction and switching function. Due to the non-relativistic internal dynamics of A, B this overlap will generally be non-zero, i.e., there is no valid choice of the states of A, B such that the upper bound vanishes, and in fact, there is always some signalling. This is a major difference with respect to the previous example of the two-level system, where one could choose valid states such that $\mathcal{S}=0$ (e.g. for $\langle \hat{\mu}_\textsc{a}\rangle=0$). In this sense, the influence of the tails is more pernicious in the Unruh-Wald model than it is for the usual Unruh-DeWitt detector.

\subsection{Signaling between a spin and a continuous pointer variable} \label{point}

Here we consider a setup similar to the thought experiments in \cite{PhysRevD.106.076018} which investigate the interplay between complementarity and relativistic causality. The `sender' is a spin system locally coupled to the field in a spacetime region and the `receiver' is a continuous pointer variable linearly coupled to the field in another spacetime region. Namely,
\begin{align}
    &\hat{J}_{\textsc{a}}(\bm{x},t )= \chi_{\textsc{a}}(t) F_{\textsc{a}}(\bm x) \hat{\sigma}_{\textsc{a}}(t) \\
    &\hat{J}_{\textsc{b}}(\bm{x},t )= \chi_{\textsc{b}}(t) F_{\textsc{b}}(\bm x) \hat{k}_{\textsc{b}}(t)
\end{align}
in the Hamiltonian densities~\eqref{haJ} and \eqref{hbJ}, where $\hat \sigma_\textsc{a}$ is the internal degree of freedom of the sender that couples to the field and $\hat{k}_{\textsc{b}}$ is the conjugate to the pointer variable of the receiver $\hat{x}_{\textsc{b}}$, i.e., $[\hat{x}_{\textsc{b}},\hat{k}_{\textsc{b}}]=\ii \hbar\openone$. We consider for simplicity, that the pointer variable does not have internal dynamics,i.e. $\hat{k}_{\textsc{b}}(t)= \hat{k}_{\textsc{b}}$, so that it only `shifts' based on its interaction with the field and the influence of A. Taking into account the definitions 
\eqref{y} and \eqref{s}
\begin{align}\label{gvar}
    \mathcal{S}=\text{Var}\;G_{\textsc{r}}[ \hat J_\textsc{b},\braket{\hat J_\textsc{a}}],
\end{align}
and that in this case
\begin{align}
    G_{\textsc{r}}[ \hat J_\textsc{b},\braket{\hat J_\textsc{a}}]=\hat{k}_{\textsc{b}} G_{\textsc{r}}[ \chi_{\textsc{b}}F_{\textsc{b}}\hat J_\textsc{b},\braket{\hat J_\textsc{a}}],
\end{align}
we get that
\begin{equation}
\mathcal{S} = \left(\int \d \bm x \d t \; \chi_{\textsc{b}}(t) F_{\textsc{b}}(\bm{x})G_{\textsc{r}}[\braket{\hat J_\textsc{a}}]( \bm x,t )   \right)^2 (\Delta k_{\textsc{b}})^2 \label{deltak}
\end{equation}
where \mbox{$(\Delta k_{\textsc{b}})^2= \langle \hat{k}^2_{\textsc{b}}\rangle-\langle \hat{k}_{\textsc{b}}\rangle^2$} and assuming for convenience that the conjugate pointer variable is centered around zero ($\langle \hat{k}_{\textsc{b}}\rangle=0$). We see that the initial variance of the conjugate pointer variable modulates the amount of signaling from A to B, times the integral prefactor in \eqref{deltak} that quantifies the causal overlap of the sender and the receiver. Notice that if the pointer variable conjugate momentum starts in an eigenstate of $\hat k_\textsc{b}$, then $\Delta k_{\textsc{b}}=0$ and there is no signalling. This is expected since in that case the state of the receiver is an eigenstate of the local interaction Hamiltonian and there is no internal dynamics, so time evolution on the partial system of the receiver is trivial.

Notice that if we were to consider that sender is coupled to the field for all times (adiabatic switching) even if the sender and receiver are fixed at two distant spatial locations $\bm{x}_{\textsc{a}}, \bm{x}_{\textsc{b}}$ they  are not spacelike separated if the interaction of A is always `on'. In this case, the prefactor of \eqref{deltak} is given by 
\begin{equation}
    \int \d t \d t' \chi_{\textsc{b}}(t)  G^{\textsc{ab}}_{\textsc{r}}(t,t') \chi_{\textsc{a}}(t')\langle \hat{\sigma}_{\textsc{a}}(t') \rangle, \label{pointlike}
\end{equation}
where \mbox{$G^{\textsc{ab}}_{\textsc{r}}(t,t') =   G_{\textsc{r}}(t,\bm{x}_{\textsc{b}}, t',\bm{x}_{\textsc{a}})$}. 
 In \cite{PhysRevD.106.076018} it was shown that the distinguishability of B (i.e. the ability of B to distinguish between the up and down state of the spin A) depends on the overlap $ \int \d t \d t' \chi_{\textsc{b}}(t)  G^{\textsc{ab}}_{\textsc{r}}(t,t') \chi_{\textsc{a}}(t')$. Comparing with the prefactor \eqref{pointlike} that goes into our signaling estimator, we see that the distinguishability ignores the effect of the internal dynamics of the spin and the choice of initial state.  This suggests that there might be cases where the ability of A to signal to B is different than e.g. the distinguishability. 
Thus, it would be interesting to investigate the effect of the internal dynamics and choice of initial state on the causality of the setup.
As we will see explicitly in subsection~\ref{betweentwolevels}, considering the internal dynamics of the sender and receiver is fundamental to deciding if  localized (but not compactly supported) systems can or cannot signal to each other (even with strongly decaying tails).

\subsection{Signalling between quantum field probes}
Finally, we can consider signalling between two quantum fields that act as probes for a third quantum field, as in \cite{fewster2020quantum}. In this case, one can consider a linear coupling given by detector operators of the form
    \begin{align}
     \hat J_\mu=\Lambda_\mu\hat{\Psi}_\mu
\end{align}
where $\hat{\Psi}_{\textsc{a,b}}$ are quantum fields defined on the same spacetime background, possibly with different dynamics from the one from the mediating field $\hat\phi$, and $\Lambda_{\textsc{a,b}}$ are the coupling functions associated with each interaction. 

In this case, the signalling estimator will be the variance of the following operator
\begin{align}
    \hat{\Sigma}=\int \d V\Lambda_{\textsc{b}}(\mathsf{x}) \hat\Psi_\textsc{b}(\mathsf{x})G_\textsc{r}[\Lambda_{\textsc{a}}\braket{\hat\Psi_\textsc{a}}](\mathsf{x}).
\end{align}
This variance, and therefore the signalling estimator, is given by
\begin{align}
  \nn&  \mathcal{S} =\iint \d V\d V'\Gamma(\mathsf{x}) \Gamma(\mathsf{x}')W(\mathsf{x},\mathsf{x}')\\
  &-\left(G_\textsc{r}[\Lambda_{\textsc{b}}\braket{\hat\Psi_\textsc{b}},\Lambda_{\textsc{a}}\braket{\hat\Psi_\textsc{a}}]\right)^2
\end{align}
where we have defined the function
\begin{align}
   \Gamma (\mathsf{x})=\Lambda_{\textsc{b}}(\mathsf{x})  G_\textsc{r}[\Lambda_{\textsc{a}}\braket{\hat\Psi_\textsc{a}}](\mathsf{x}),
\end{align}
and $W(\mathsf{x},\mathsf{x}')$ is the Wightman function of the probe B in its initial state, that is
\begin{align}
    W(\mathsf{x},\mathsf{x}')=\braket{\hat\Psi_\textsc{b}(\mathsf{x})\hat\Psi_\textsc{b}(\mathsf{x'})}.
\end{align}

There are several aspects of this expression that merit at least some qualitative discussion. Note that if probe B is prepared in a coherent state (of a linear field), then the two-point function takes the form
\begin{align}
    W(\mathsf{x},\mathsf{x}')= W_0(\mathsf{x},\mathsf{x}')+\braket{\hat\Psi_\textsc{b}(\mathsf{x})}\braket{\hat\Psi_\textsc{b}(\mathsf{x'})}
\end{align}
where $W_0(\mathsf{x},\mathsf{x}')$ is the vacuum two-point function. Then for any coherent state of detector B, the signalling received from A (at leading order) is identical and is given by 
\begin{align}
    \mathcal{S} =\iint \d V\d V'\Gamma(\mathsf{x}) \Gamma(\mathsf{x}')W_0(\mathsf{x},\mathsf{x}').
\end{align}
The intuition behind this is that linear fields obey the superposition principle, in such a way that if the field B is prepared in a coherent state the fluctuations around the mean value are the same as in the vacuum state.

Second, one could be alarmed when observing that, given that it is known that for thermal states the self correlations, given by the Wightman function, increase with the temperature, so would the signalling estimator.  This would lead one to conclude, mistakenly, that signalling is somehow aided by thermal fluctuations. In that case, however, the signalling estimator is just a loose bound given that thermal states are mixed, and our signalling estimator does not capture the efficiency of the signalling process as well as in pure states, for which our signalling estimator gives exactly the quantum Fisher information at leading order. This observation exemplifies the fact that one cannot use this signalling estimator for comparing signalling in situations where a mixed state is prepared with situations where a pure state is prepared.

\section{Signalling estimator for smeared UDW detectors: gap dependence and resonant phenomena}\label{betweentwolevels}

In this section we will analyze of the signalling estimator for the more familiar case of the usual smeared Unruh-DeWitt detector~\eqref{h1}. Namely, the interaction Hamiltonian of two inertial comoving two-level systems interacting with a scalar field is given by
\begin{align}
   \hat H_{\text{sm}} =\sum_{\nu=\textsc{a},\textsc{b}}\lambda_{\nu} \chi_{\nu}(t) \hat{\mu}_{\nu}(t)\otimes \int \d^n \bm x\, F_{\nu}(\bm x)\hat{\phi}(t,\bm x),
\end{align}
where 
 $\hat{\mu}_{\nu}(t)=\alpha^\frac{3-n}{2}(e^{\ii\Omega_{\nu} t}\hat{\sigma}_{\nu}^++e^{-\ii\Omega_{\nu} t}\hat{\sigma}_{\nu}^-)$,  ${\sigma}_{\nu}^+,{\sigma}_{\nu}^-$ are the operators that take the ground state to the excited state, and $\Omega_{\nu}$ is the energy gap between the two levels, and $\alpha$ is a scale with dimensions of energy so that the coupling constant $\lambda_\nu$ is dimensionless\footnote{The reason we enforce the coupling strength to be dimensionless is so that the Fisher information is dimensionless. In 3+1 spacetime dimensions $\alpha$ is dimensionless and can be set to 1 for our purposes. For other spacetime dimensions $\alpha$ will be given by one of the problem's reference scales.}.
We wish to specify which features of the interaction are relevant to determine the signalling estimator \eqref{y}. 

 To particularize the general expression for the signalling estimator~\eqref{signUDWtotalfreq} to the two-level Unruh-DeWitt detector in flat space-time of $n+1$ dimensions, we make the following identification 
\begin{align}
\hat J_{\nu}(\mathsf{x})=\chi_{_{\nu}}(t)F_{_{\nu}}(\bm{x})\hat{\mu}_{_{\nu}}(t) . 
\end{align}
 where $\nu=\text{A},\text{B}$. In flat spacetime, the field propagators are translationally invariant, meaning that they only depend on the difference of spacetime points. This allows us to write the expression for the functions $I(\Omega,\Omega')$ appearing in the expression for the optimal signalling estimator defined in \eqref{signUDWtotalfreq} as
\begin{align}\label{laIesa}
 \nn&I(\Omega,\Omega')=\int \d t \d \bm{x}\, G_\textsc{r}(t,\bm x)\int \d \bm y F_\textsc{a}(\bm y) F_\textsc{b}(\bm x+\bm{y})\\
&\times\int\d s\chi_\textsc{a}(s)e^{\ii\Omega s} 
  \chi_\textsc{b}(t+s)e^{\ii\Omega' (t+s)},
\end{align}
which is derived by simply substituting $G_\textsc{r}(\mathsf x,\mathsf x')=G_\textsc{r}(\mathsf x-\mathsf x')$ and performing a change of variables.

Further, the propagators in flat spacetime are tempered distributions, which in particular implies that they admit a Fourier transform. In terms of the Fourier transform the integral takes the form 
\begin{align}\label{intfourier}
 \nn&I(\Omega,\Omega')=\int \d \bm k \d k^{0}\, \tilde{G}_\textsc{r}(\bm k,k_0) \tilde F_\textsc{b}^{*}(\bm k)\tilde F_\textsc{a}(\bm k)\\
&\times\tilde{\chi}_\textsc{b}^{*}(k_0+\Omega ')\tilde{\chi}_\textsc{a}(k_0+\Omega),
\end{align}
where $\tilde F_\textsc{a,b}$ is the $n$-dimensional Fourier transforms of $F_\textsc{a,b}$ and $\tilde \chi_\textsc{a,b}$ is the one-dimensional Fourier transforms of $\chi_\textsc{a,b}$:
\begin{align}
    \tilde{F}(\bm k)=\frac{1}{(2\pi)^{n/2}}\int \d^n \bm x e^{-\ii \bm k \bm x} F(\bm x)\\
     \tilde{\chi}( k_0)=\frac{1}{(2\pi)^{1/2}}\int \d \bm t e^{-\ii k_0 t} \chi(t).
\end{align}
For the Klein-Gordon field the Fourier transform of the retarded propagator is given by
\begin{align}
    \tilde{G}_\textsc{r}(\bm k,k^0)=\frac{1}{(2\pi)^{(n+1)/2}}\frac{1}{-(k^{0}-\ii\epsilon)^2+\omega^{2}_{\bm k}}
\end{align}
where $\omega^2_{\bm k}=m^2+\bm k^2$ and $\epsilon$ is a regulator that will have to be taken to go to zero after the integrals are performed.
Therefore, the integral in \eqref{intfourier} is given by
\begin{align}
 \nn&I(\Omega,\Omega')=\frac{1}{(2\pi)^{n+1}}\lim_{\epsilon\to0^+}\int \d \bm k \,  \tilde F_\textsc{b}^{*}(\bm k)\tilde F_\textsc{a}(\bm k)\\
&\times\int\d k_0 \frac{\tilde{\chi}_\textsc{b}^{*}(k^0+\Omega ')\tilde{\chi}_\textsc{a}(k^0+\Omega)}{-(k^{0}-\ii\epsilon)^2+\omega^{2}_{\bm k}}. \label{iota}
\end{align}

We are interested in analyzing a) the dependence of the estimator on the frequencies $\Omega_\textsc{a},\Omega_\textsc{b}$, and b) the behavior of the estimator as we vary the smearing and switching functions of the detectors.


Within this section we will study the dependence of the signalling estimator in the particular case of  two UDW detectors that interchange signals through a massless scalar field in a  flat spacetime of 3+1 dimensions. In order to simplify the explicit evaluation we recall that the two detectors are comoving and at rest in some inertial frame, and  further assume that  The switching and smearing functions are given by Gaussian functions. The centers of the Gaussians that define the detectors' spacetime smearing will be separated by a constant spacetime vector, which in the comoving coordinate frame takes the explicit components $\mathsf z=(z_0, \bm{z})$. We call $z_0:=\Delta$ the time distance and $|\bm{z}|:=L$ the spatial distance. Both smearing functions have spatial width $R$ and temporal width $T$ (see Eqs.~\eqref{smearing} and \eqref{switching}). 

Since we are in 3+1 dimensions we take $\alpha=1$ and the signaling estimator is dimensionless. Recall that the maximum signaling between the two detectors (equation \eqref{MaxSignalling}) is given by
\begin{align}
\mathcal{S}_{\text{max}}=\left(|I(\Omega_{\textsc{a}},-\Omega_{\textsc{b}})|+ |I(\Omega_{\textsc{a}},\Omega_{\textsc{b}})|\right)^2.
\end{align}
Through simple analysis (see Appendix~\ref{APPENDIX2}), particularizing the Green function in expression \eqref{laIesa} to the case of a massless scalar field in 3+1 dimensions (i.e.,~\eqref{lienardwiechert}) we get 
\begin{align}
I(\Omega_A, \Omega_B)= \frac{R}{\sqrt{\pi}L} e^{-\frac{L^2}{4R^2}}\int_0^{\infty} \d u \; u & e^{\frac{-u^2}{4}}(e^{\frac{uL}{2R}}-e^{-\frac{uL}{2R}}) \nn \\
& \times I_p(Ru,\Omega_A, \Omega_B)
\end{align}
where 
\begin{align}\label{iotap}
  I_p(L,\Omega_A, \Omega_B)=  -\frac{T}{8\pi^{3/2}L} e^{-\frac{T^2(\Omega_B-\Omega_A)^2}{4}} e^{-\frac{(L+\Delta)^2}{4T^2}} \nn \\
  \times e^{\ii (\Omega_B+\Omega_A)L}
\end{align}
 happens to be the value of $I(\Omega_A, \Omega_B)$ for two pointlike detectors with identical switching separated by a spatial distance $L$. In other words, we can express $I(\Omega_A, \Omega_B)$ as a spatial integral of the expression for pointlike detectors.

Regarding the behavior of the signalling estimator with respect to the detectors's energy gap, in general, we would  expect  that signalling decreases in the limit of very large gap of either detector. The fact that the signalling estimator vanishes in the large gap limit is clear from equation~\eqref{iota}, since in this case the signalling estimator is given by an integral of smooth functions and the dependence on the internal frequencies of the detectors is just a translation in momentum space. For smooth, integrable smearing profiles and switching functions the limit of the integral when $\Omega_{\textsc{a}}$ or $\Omega_{\textsc{b}}$ goes to infinity can be taken inside the integral, and this limit vanishes. We would also expect the signalling estimator to be the largest in resonance (that is, when $|\Omega_{\textsc{a}}|=|\Omega_{\textsc{b}}|$).

\begin{figure}
\centering
\includegraphics[width=0.47\textwidth]{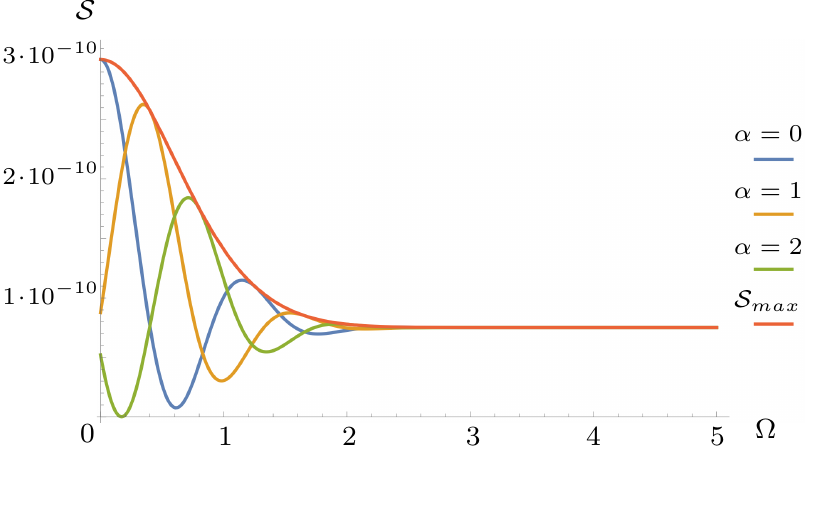}
\caption{ The three oscillatory curves (blue, orange, green) give the signaling estimator for different states of detector A as parametrised by a phase $\alpha$ ($\alpha=0,1,2$) as a function of the detectors' gap $\Omega_\textsc{a}=\Omega_\textsc{b}=\Omega$. The red curve that is decreasing monotonically is the $ \mathcal{S}_{\text{max}}$. This is for pointlike detectors that are separated by $L$ and both switchings are picked around zero with temporal width $T$, and such that $L/T=5$. Notice that the signalling estimator at this spatial separation is $\mathcal{S}\ll 1$, so the detectors are effectively out of causal contact.}
\label{fig1}
\end{figure}

It is interesting to consider the pointlike limit. In this case (see Figure \ref{fig1}) the signalling does not vanish when $\Omega\to\infty$ while in resonance $|\Omega_{\textsc{a}}|=|\Omega_{\textsc{b}}|=\Omega$. This is expected since in the pointlike limit the  smearing function becomes a delta which is not a smooth function and the arguments given above do not apply. Furthermore, in the pointlike limit, the signalling estimator reaches an asymptote as the resonant frequency $\Omega$ grows (with every other parameter fixed) independently of the initial state of detector A if the state is of the form $\hat\rho_\textsc{a}=\frac{1}{2}(\openone + \cos{\alpha}\hat\sigma_x+\sin{\alpha}\hat\sigma_y)$. The value of the upper bound for the signalling estimator~\eqref{MaxSignalling} behaves, as expected, as an enveloping curve for different values of the initial state of A. Whereas the behavior of the maximum signalling is monotonic, the behavior of the particular signalling estimator for a fixed state of detector A (Eq. ~\eqref{signUDWtotalfreq}) becomes oscillating. We relate this phenomenon, already observed in~\cite{PhysRevD.106.045014}, with the fact that for a fixed state and an interaction duration long enough, the detector interacts mostly with some frequencies of the field around its internal frequency $\Omega_\textsc{a}$. Therefore, if the detectors are separated by a fixed spacial distance, it is to be expected that signalling will be maximum when they are separated by one of these wavelengths. This can be seen directly from equation~\eqref{iotap}, where we see that the signalling estimator will oscillate in $\Omega$ exactly with frequency $L^{-1}$.

\begin{figure}
\centering
\includegraphics[width=0.5\textwidth]{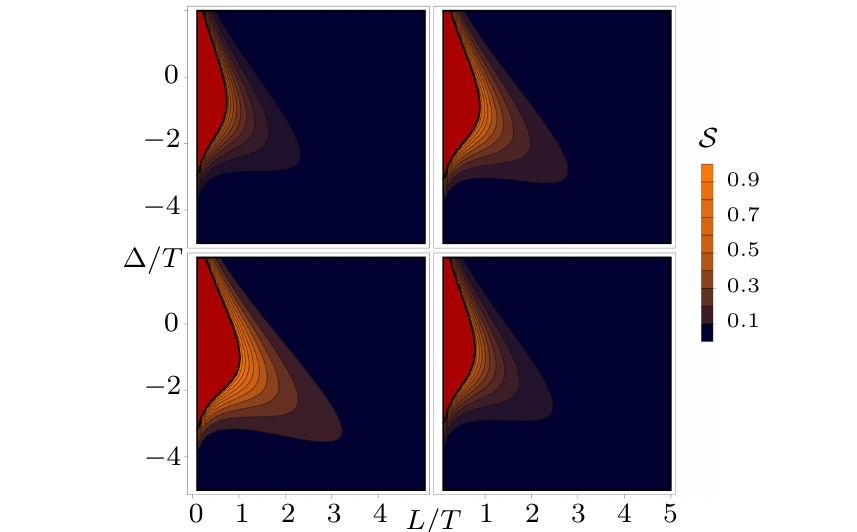}
\caption{ Signalling estimator between two pointlike detectors that are separated by a spatial distance $L$ and with a time lapse $\Delta$. The picks of the switchings are separated by time $\Delta$ with temporal width $T$. Detector B (the receiver) has internal gap $\Omega_{\textsc{b}}=2$ and is centered around zero. The color bar quantifies how much signal it can receive from A depending on where A is in space and time, for different values of $\Omega_{\textsc{a}}$($\Omega_{\textsc{a}}=1,2,3,4$ from up left to down right).}
\label{fig2}
\end{figure}

We also observe a resonance phenomenon in Figure~\ref{fig2} where both detectors are pointlike separated by spatial distance $L$ and both switchings are Gaussian functions of width $T$. Detector B (the receiver) is centered around zero. As expected, we see that the signaling is higher when A is centered on the (smeared) past light cone of B, and when the two frequencies match. 

\begin{figure}
\centering
\includegraphics[width=0.5\textwidth]{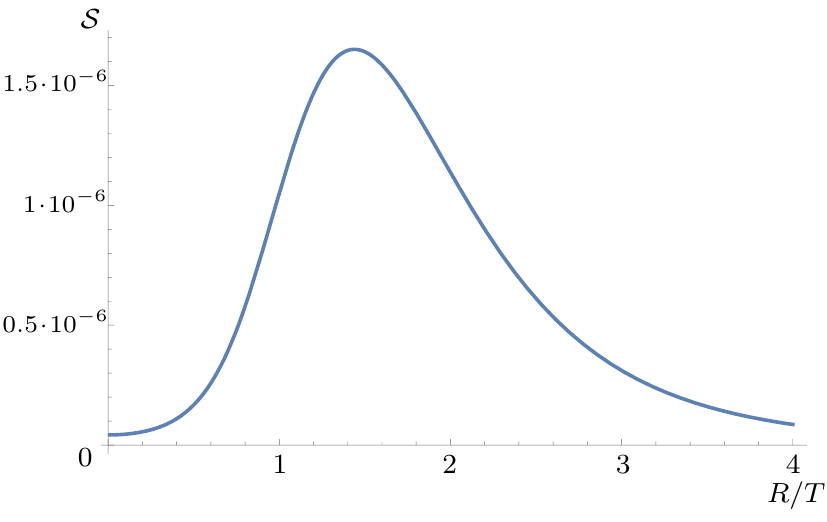}
\caption{ Signalling estimator for two smeared detectors with spatial smearings of width $R$ and whose centers are separated by a spatial distance $L$. The spatial separation $L$ is such that $L/T=4$ (where $T$ is the duration of the interaction). Both switching functions are peaked around $t=0$, so the two detectors are effectively spacelike separated ($\mathcal{S}\ll 1$). We have set $\Omega_\textsc{a}=\Omega_\textsc{b}=\Omega$.}
\label{fig3}
\end{figure}

Finally, in Figure~\ref{fig3} we see the effect of the spatial smearing on signaling when the centers of the two detector-field interactions are in spacelike separation. To see how much the Gaussian smearing affects signalling and when we stop being in effective spacelike separation as we increase the smearing width $R$, we study the worst-case scenario: in order to maximize the signalling estimator, we choose the frequencies such that the detectors are in resonance, and we plot the signaling as a function of the spatial width $R$ of the smearing for constant $L/T=4$. One would expect that the bigger the width the causal connection between the tails of the two smearings grows which would lead to more signaling. Indeed, this is what happens until some critical value, after which the signaling decreases (see figure \ref{fig3}). This might seem counter-intuitive, but since the smearing function is a normalized density, the interaction overall vanishes in the limit of $R/T\rightarrow \infty$. In this sense, even though we would expect that the tails are indeed enabling signalling between the detectors, we also observe a `dilution' effect. Similar behavior was also observed previously in \cite{PhysRevD.92.104019}.

\section{Exponential bounds on signalling for UDW detectors in flat spacetimes}\label{sec:bounds}

In this section we will take advantage of the analytic structure of the retarded propagators in flat spacetime for deriving frequency-independent, exponential bounds on the signalling estimator between two UDW detectors. 

\subsection{Bounds for smooth spacetime smearings}

To analyze the dependence of signalling on the switching and smearing functions it is technically useful to make some assumptions. Let us recover the spacetime smearing notation
\begin{align}\label{89fornow}
    \Lambda_{\textsc{a,b}}(\mathsf{x})=F_{\textsc{a,b}}(\bm x)\chi_{\textsc{a,b}}(t). 
\end{align}

Recall that, conventionally, $\chi_{\textsc{a,b}}(t)$ is considered to be a dimensionless function that just modulates the coupling strength in time\footnote{This is the most common convention although sometimes nascent delta switching functions are employed.}, unlike the smearing function which is a density over space with dimensions of $[\text{length}]^{-n}$. First, let us assume that the smearings and switchings of the detectors A and B only differ by a spacetime translation of ($n+1$)-vector $\mathsf{z}$. Without loss of generality we set \mbox{$\Lambda_\textsc{b}(\mathsf{x})=\Lambda(\mathsf{x})$}, $\Lambda_\textsc{a}(\mathsf{x})=\Lambda(\mathsf{x}-\mathsf{z})$,
where $\Lambda$ is a function localized\footnote{We will clarify more rigorously what we mean with the word localized later on, but for now one can think of a function that decays  fast enough far away from the origin.} at around zero.

Second, we consider that the spacetime decays with some given dimensional length and timescales. We can then write the spacetime smearing in terms of a dimensionless smearing function $\Xi$ as follows 
\begin{align}
     \Lambda(\mathsf{x})=\frac{{T}}{|L|}\Xi(L^{-1}\mathsf{x}) \label{xi}
\end{align}
where $L$ is a strictly positive matrix whose entries have  units of length, $|L|$ is its determinant and ${T}$ is a ``duration'' timescale associated with the interaction. Since we assumed by construction that in the the coordinate frame $(t,\bm x)$ the spacetime smearing is a product of the temporal and spatial smearing (Eq.~\eqref{89fornow}) the matrix $L$ in those coordinates has the form
\begin{align}
    L=\begin{pmatrix}
    {T}&\bm 0\\
    \bm 0 & R\openone_n
    \end{pmatrix}.
\end{align}
This means that $|L|={T} R^n$ and we can rewrite \eqref{xi} as \mbox{$ \Lambda(\mathsf{x})=\frac{1}{R^n}\Xi(L^{-1}\mathsf{x})$}. In this situation ${T}$ represents the time scale associated with the switching and $R$ the spatial scale associated with the smearing. 

We say that the dimensionless function $\Xi$ is exponentially localized at zero if
\begin{align}
    \Xi(\mathsf{y})e^{L\upeta \mathsf y}\in L^{2}(\mathbb{R}^n)
\end{align}
where $\upeta \mathsf y$ represents the contraction of the $(n+1)$-vector $\mathsf{y}$ and the covector $\upeta$, $\mathsf{y}= L^{-1} \mathsf{x}$ for all covectors $L\upeta$ with $\norm{L\upeta}_{e}\leq1$, where $\norm{\cdot}_e$ is the Euclidian norm in the reference frame with coordinates $y^{0},\bm y$. This notion of localization, of course, depends on the reference frame. However, given that there is a preferred class of reference frames, which is the one where spacetime smearings factorize as a function of time and a function of the spacial coordinates, it is natural to demand exponential localization in these frames.
Note that if we define $u^\mu=(1,\bm 0)$ in coordinates $(t,\bm x)$, then the Euclidian norm can be written covariantly as $\norm{\mathsf{y}}_{e}=\sqrt{y^\mu y_\mu+2(y^{\mu} u_{\mu})^2}$.

Finally, we will assume that 
 \begin{align}
     \frac{\norm{\Xi e^{L\upeta \cdot}}_2}{\norm{\Xi}_2}=\mathcal{O}(1),
 \end{align}
 for $\norm{L\upeta}_{e}\leq1$, where the notation $ e^{L\upeta \cdot}$ represents a function over spacetime vectors and the dot represents its argument\footnote{Basically, $f=e^{L\upeta \cdot}$ means that $f(\mathsf{z})= e^{L\upeta \mathsf{z}}$.}.  
This condition represents mathematically the idea that the function takes values significantly different from zero only in a region around the unit ball in dimensionless coordinates.

An example of such a function would be a Gaussian:
\begin{align}
     \Xi(\mathsf{y})=\frac{1}{(2\pi)^{n+1/2}}e^{-\frac{\norm{\mathsf{y}}^2}{2}},
\end{align}
 \begin{align}
     \frac{\norm{\Xi e^{L\upeta \cdot}}_2}{\norm{\Xi}_2}=e^{\frac{\norm{L\upeta}^2}{2}}\leq e^{\frac{1}{2}}.
 \end{align}

The expression for the integral in \eqref{intfourier} now takes the covariant form
\begin{align}\label{intfourier2}
 \nn&I(\Omega,\Omega')\\
 &={T}^2 \int \d ^{n+1}\mathsf k \, \tilde{G}_\textsc{r}(\mathsf{k}) \tilde{\Xi}^*(L\mathsf{k}+\Omega L\mathsf{u})\tilde{\Xi}(L\mathsf{k}+\Omega' L\mathsf{u})e^{-\ii\mathsf{k} \mathsf{z}}.
\end{align}
where we have defined the dimensionless Fourier transform
\begin{align}
    \tilde{\Xi}(\upkappa)=\frac{1}{(2\pi)^{(n+1)/2}}\int\d^{n+1} \mathsf{y}e^{-\ii\upkappa\mathsf y}\Xi(\mathsf{y}).
\end{align}

We are interested in the behavior of this expression as the separation between the center of the spacetime smearings, $\mathsf{z}$, varies. We will use analytic properties of the function $\Xi$ and the propagator $G_{\textsc{r}}$.
On the one hand, the exponential localization of $\Xi$ implies that $\tilde{\Xi}$ admits an analytic extension, $\tilde{\Xi}(\upkappa-\ii L\upeta)$ with $\norm{L\upeta}_e\leq 1$. This is a classic result that can be easily checked from the expression 
\begin{align}
    \nn\tilde{\Xi}(\upkappa-\ii L\upeta)&=\frac{1}{(2\pi)^{(n+1)/2}}\int\d^{n+1} \mathsf{y}e^{-\ii(\upkappa-\ii L\upeta)\mathsf y}\Xi(\mathsf{y})\\
    &=\frac{1}{(2\pi)^{(n+1)/2}}\int\d^{n+1} \mathsf{y}e^{-\ii\upkappa \mathsf y}e^{-L\upeta\mathsf y}\Xi(\mathsf{y}).
\end{align}
Since $e^{-L\upeta\mathsf y}\Xi(\mathsf{y})$ is square-integrable, its Fourier transform exists and it is also square-integrable. This Fourier transform then gives the unique analytic extension of $\tilde{\Xi}$.

On the other hand, by a known result in distribution theory (see \cite{MR0493420}) the Fourier transform of a distribution whose support is contained in a cone $\mathsf{V}$ is the boundary value of an analytic function in the complex region \mbox{$\mathbb{R}^n-\ii\mathsf{V}^*$}, where $\mathsf{V}^*$ is the dual cone. Further, this function is also polynomially bounded in $\mathsf k$. This means in particular that $\tilde{G}_\textsc{r}(\mathsf{k})$ can be analytically extended to $\tilde{G}_\textsc{r}(\mathsf{k}-\ii \upeta)$ where $\upeta$ is an arbitrary future-oriented timelike vector\footnote{In QFT it is common to consider the analytic extension of the distribution $\tilde{G}_\textsc{r}$ as a way of defining the action of the distribution over arbitrary smooth functions. In this sense, the covector $\upeta$ is considered to be a small regulator. However, this is not what we are doing here. Recall that we use the analytic extension of both the propagator and the smearing to the complex plane in order to establish bounds for the signalling estimator, thus for us the magnitude of $\upeta$ will not be a small value.}.

In our case, we are considering that the distribution $\tilde{G}_\textsc{r}$ acts over test functions that also admit an analytic extension. This means that one can define the action of the distribution over the functions by shifting the contour of integration in  \eqref{intfourier2}:
\begin{align}\label{intfourier3}
 \nn& I(\Omega,\Omega')
={T}^2 e^{-\upeta\mathsf{z}}\int \d ^{n+1}\mathsf k \, \tilde{G}_\textsc{r}(\mathsf{k}-\ii\upeta) e^{-\ii\mathsf{k} \mathsf{z}}\\
&\times \tilde{\Xi}^*(L\mathsf{k}+\Omega L\mathsf{u}-\ii L\upeta)\tilde{\Xi}(L\mathsf{k}+\Omega' L\mathsf{u}-\ii L\upeta),
\end{align}
for any timelike covector $\upeta$ such that $\norm{L\upeta}_e\leq 1$. The value of the integral is independent of $\upeta$. However, for any contour where $\upeta$ is not the zero covector, the value of $I(\Omega,\Omega')$ is given by an integral of a smooth, complex-valued function. Therefore, one can apply bounds to the modulus of the integral in the usual way.
Indeed, by taking the modulus of the integral one arrives at the following bound:
\begin{align}\label{intfourierbound}
 \nn& |I(\Omega,\Omega')|
\leq {T}^2 e^{-\upeta\mathsf{z}}\int \d ^{n+1}\mathsf k \, |\tilde{G}_\textsc{r}(\mathsf{k}-\ii\upeta)|\\
&\times |\tilde{\Xi}^*(L\mathsf{k}+\Omega L\mathsf{u}-\ii L\upeta)\tilde{\Xi}(L\mathsf{k}+\Omega' L\mathsf{u}-\ii L\upeta)|.
\end{align}
 Recall that the value of the integral is independent of $\upeta$, and it is well defined even when $\upeta$ is the zero covector. However, the bound is not independent of $\upeta$: every value of $\upeta$ provides a different bound. We will choose particular values of $\upeta$ that lead to exponentially decaying bounds with the distance and the time lapse between the interactions.

The bound given by \eqref{intfourierbound} is valid for any propagator that has support only in the future lightcone. In order to provide a bound which is independent of the detector gaps $\{\Omega_\textsc{a},\Omega_\textsc{b}\}$, we need to use properties of Klein-Gordon field propagator.

In covariant form, the expression for the analytic continuation of the retarded propagator is given by
\begin{align}
    \tilde{G}_\textsc{r}(\mathsf{k}-\ii\upeta)=\frac{1}{(2\pi)^{(n+1)/2}}\frac{1}{(\mathsf{k}-\ii\upeta)^2+m^2}.
\end{align}
This is a regular function of $\mathsf{k}$ whose modulus squared is
\begin{align}
    |\tilde{G}_\textsc{r}(\mathsf{k}-\ii\upeta)|^2=\frac{1}{(2\pi)^{(n+1)/2}}\frac{1}{(\mathsf{k}^2+m^2-\upeta^2)^2+4(\mathsf{k}\upeta)^2}.
\end{align}

This expression has a global maximum dependent on $\upeta$ and $m$. When $|\upeta^2|\geq m^2$, the maximum is given by
\begin{align}
   \nn\text{max} \left( |\tilde{G}_\textsc{r}(\mathsf{k}-\ii\upeta)|\right)&= \frac{1}{(2\pi)^{(n+1)/2}}\frac{1}{|\upeta^2|+m^2}\\
  & \leq\frac{1}{(2\pi)^{(n+1)/2}}\frac{1}{|\upeta^2|}, \label{b1}
\end{align}
 and when $|\upeta^2|\leq m^2$
\begin{align}
    {\text{max}} \left(|\tilde{G}_\textsc{r}(\mathsf{k}-\ii\upeta)|\right)= \frac{1}{(2\pi)^{(n+1)/2}}\frac{1}{2m\sqrt{|\upeta^2|}}. \label{b2}
\end{align}

Under our assumptions about exponential localization the factors dependent on $\Xi$ are just form factors that are mostly independent of $\upeta$ provided that $\norm{L\upeta}_e\leq1$. Therefore, we can bound the integral as (see Appendix~\ref{app:INTBound})
\begin{align}\label{intfourierbound6}
  |I(\Omega,\Omega')|
\leq C\frac{{T} e^{-\upeta\mathsf{z}}}{R^n(|\upeta^2|)},\quad \text{when }|\upeta^2|> m^2
\end{align}
and
\begin{align}\label{intfourierbound7}
  |I(\Omega,\Omega')|
\leq C\frac{{T} e^{-\upeta\mathsf{z}}}{R^n2m\sqrt{|\upeta^2|}},\quad \text{when }|\upeta^2|\leq m^2,
\end{align}
for a dimensionless constant $C\approx 1$ independent of $\upeta$, together with the constrain $\norm{L\upeta}_e\leq1$.

Therefore, the optimal signalling estimator in \eqref{signUDWtotalfreq} can be bounded by
\begin{align}\label{masslesscase}
       \mathcal{S}_{\text{max}}\leq\frac{4C^2{T}^2 e^{-2\upeta\mathsf{z}}}{R^{2n}\upeta^4},\quad \text{when }|\upeta^2|> m^2,
\end{align}
and
\begin{align}
       \mathcal{S}_{\text{max}}\leq\frac{C^2 {T}^2 e^{-2\upeta\mathsf{z}}}{R^{2n} m^2|\upeta^2|} ,\quad \text{when }|\upeta^2|\leq m^2.
\end{align}

Recall that these bounds are valid for all timelike $\upeta$ such that $\norm{L\upeta}_e\leq1$. Whether the bound is useful or not will depend on whether the exponential is a decreasing exponential or not. Indeed, if the exponent is positive, e.g. if $\upeta\propto\mathsf{z}$ and $\mathsf{z}$ is in the future lightcone, the bound just implies that the signalling estimator is less than infinity as $\mathsf{z}$ grows in a timelike direction. 

Recall that if $\mathsf{z}$ is a past timelike ($n+1$)-vector, detector A is mostly localized in the past of B, whereas if $\mathsf{z}$ is spacelike, detector A is mostly localized in spacelike separation from B.

Consider first the massless case, given by the bound in equation \eqref{masslesscase}.
If $\mathsf{z}$ is past-timelike, then a natural choice for $\upeta$ is given by $\upeta=\frac{1}{{T}}\mathsf{u}$, since in this case $L\upeta=\mathsf{u}$ and $\norm{L\upeta}_e=\norm{\mathsf{u}}_e=1$. In this case, $\upeta\mathsf{z}=\frac{|z^0|}{{T}}$ and the optimal signalling (which upper bounds any apparent retrocausation in the model) is given by
\begin{align}\label{retrobound}
       \mathcal{S}_{\text{max}}\leq\frac{4{T}^6C^2e^{-2\frac{|z^0|}{{T}}}}{R^{2n}}.
\end{align}
In this case there is an exponential decay in the signalling estimator as the time difference $|z^0|$ goes to infinity, with a scale given by the typical scale of the switching function ${T}$. 

In the case where $\mathsf{z}$ is spacelike, one can use the ansatz
\begin{align}
    \upeta=\left(\frac{\cos(\varphi)}{{T}},\frac{\sin(\varphi)}{R|\bm z|}\bm z\right)=\frac{\cos(\varphi)}{{T}}\mathsf{u}+\frac{\sin(\varphi)}{R}\frac{\mathsf{z}-(\mathsf{u}\mathsf{z})\mathsf{u}}{\sqrt{\mathsf{z}^2+(\mathsf{u}\mathsf{z})^2}}
\end{align}
which satisfies $\norm{L\upeta}=1$ by construction. For $\upeta$ to be timelike, one has to impose conditions over $\varphi$:
\begin{align}
    -\upeta^2=\frac{1}{{T}^2}\cos^2(\varphi)-\frac{1}{R^2}\sin^2(\varphi)>0,
\end{align}
which implies
\begin{align} \label{tanbound}
    |\tan(\varphi)|<\frac{R}{{T}}.
\end{align}
For $\upeta$ to be future directed, one imposes $\cos \varphi >0$, which implies that $\varphi\in(-\pi/2,\pi/2)$.

On the other hand, the exponent in~\eqref{masslesscase} is given by
\begin{align}
    -\upeta\mathsf{z}=\frac{\cos(\varphi)}{{T}}z_0-\frac{\sin(\varphi)}{R}|\bm z|.
\end{align}
For the exponent to be negative, the angle $\varphi$ has to further fulfill
\begin{align}
    \tan(\varphi)>\frac{R}{{T}}\frac{z_0}{|\bm z|}.
\end{align}
For every value of $\varphi$ satisfying these conditions, the maximum signalling is bounded by
\begin{align}
       \mathcal{S}_{\text{max}}\leq\frac{4{T}^6C^2}{R^{2n}}\frac{e^{2\frac{\cos(\varphi)}{{T}}z_0-2\frac{\sin(\varphi)}{R}|\bm z|}}{(\cos^2(\varphi)-\frac{{T}^2}{R^2}\sin^2(\varphi))^2}.
\end{align}
In the case where the detectors are situated in the same plane of simultaneity in the preferred reference frame, namely when $z_0=0$, the angle $\varphi$ can take any value compatible with condition \eqref{tanbound} and with $\varphi>0$, implying that $\varphi\in(0,\pi/2)$ and the decay in the signalling estimator is dominated by
  \begin{align}\label{spacelikebound}
       \mathcal{S}_{\text{max}}\leq\frac{4{T}^6C^2}{R^{2n}}\frac{e^{-2\frac{\sin(\varphi)}{R}|\bm z|}}{(\cos^2(\varphi)-\frac{{T}^2}{R^2}\sin^2(\varphi))^2}.
\end{align}
Given our bound, the scale of decay with the spatial separation of the detectors is given by $\frac{R}{2\sin(\phi)}$. If the ratio $\frac{{T}}{R}$ (the light-crossing time of the detectors size length-scale) is very large the tangent of $\varphi$ is forced to be small, implying that the bound for $\mathcal{S}_{\text{max}}$ in those regimes would cease to be helpful\footnote{In other words, the bound in~\eqref{spacelikebound} is always going to be a very conservative upper bound for $\mathcal{S}_{\text{max}}$ as the detectors approach the pointlike limit.}, as the actual decay will  be faster. If ${T}\sim R$, one can choose e.g. $\varphi=\frac{\pi}{3}$, which gives the bound
 \begin{align}
       \mathcal{S}_{\text{max}}\leq 4C^2\frac{e^{-\frac{|\bm z|}{R}}}{R^{2(n-3)}}.
\end{align}
which shows that signalling indeed decays exponentially with the separation of the detectors.

\subsection{The pointlike limit}
As it was noted in last subsection, the signaling bounds as defined by equations \eqref{retrobound} and \eqref{spacelikebound} do not behave well in the limit in which the time scale regulating the interactions, ${T}$, is much larger than the scale $R$ regulating the spatial profile of the detectors. This is the pointlike limit, in which one assumes that the detector is  effectively localized along a single worldline. The reason why our bounds do not behave well is that, as shown in appendix \ref{app:INTBound}, we assume that the dimensionful spacetime smearing $\Lambda$ is an $L^2$ function of the dimensionful coordinates. Clearly, in the pointlike limit, this will not be the case anymore, since the spacetime smearing will behave like a Dirac delta. 

However, it is interesting to slightly extend the analysis performed in the last section to the pointlike case. The pointlike limit plays an important role when considering relativistic causality, as shown in \cite{PhysRevD.103.085002,2020broken,PhysRevD.101.045017}. Indeed, this is the limit in which one can safely consider particle detector models without running into any problems with causality, because the full Hamiltonian density of the joint interacting system of detectors and field will commute in spacelike separation. Moreover, physically, it is important to gain some intuition as to whether the bounds found before are just too conservative, or if there is a real  problem regarding signalling because of the (in principle) non-compact support of the switching function. Let us consider two pointlike UDW detectors following static trajectories and coupled to a massless scalar field in 3+1 dimensions as a case study.

For two static detectors in flat spacetime the spacetime smearings are given by
\begin{align}
    \Lambda_{\textsc{a,b}}=\chi_{\textsc{a,b}}(t)\delta(\bm x-\bm x_{\textsc{a,b}}),
\end{align}
and following the last subsection, the maximum possible signalling is a function of the integral
\begin{align}\label{IAYAYAY}
 &I(\Omega,\Omega')\\
 \nn&=\int \d t \, G_\textsc{r}(t,\bm x_{\textsc{b}}-\bm x_{\textsc{a}})
\int\d s\chi_\textsc{a}(s)e^{\ii\Omega s} 
  \chi_\textsc{b}(t+s)e^{\ii\Omega' (t+s)}.
\end{align}
The reason why we consider the case of a massless scalar field in 3+1 is that in that case the retarded propagator admits a simple expression, since it fulfills the strong Huygens principle (it does not have any support in timelike events). Namely,
\begin{align}
    G_\textsc{r}(t,\bm x_{\textsc{b}}-\bm x_{\textsc{a}})=\frac{1}{2|\bm x_{\textsc{b}}-\bm x_{\textsc{a}}|}\delta(t-|\bm x_{\textsc{b}}-\bm x_{\textsc{a}}|).
\end{align}

Therefore, the expression \eqref{IAYAYAY} can be evaluated right away
\begin{align}
 & I(\Omega,\Omega')\\
\nn&=\frac{e^{\ii\Omega '|\bm x_{\textsc{b}}-\bm x_{\textsc{a}}|}}{2|\bm x_{\textsc{b}}-\bm x_{\textsc{a}}|} 
\int\d s\chi_\textsc{a}(s)
  \chi_\textsc{b}(|\bm x_{\textsc{b}}-\bm x_{\textsc{a}}|+s)e^{\ii(\Omega+\Omega') s}.
\end{align}
The expression is then proportional to the convolution of $\chi_{\textsc{a}}$ with $\chi_{\textsc{b}}$, and therefore has a simple expression in terms of the Fourier transforms:
\begin{align}
 & I(\Omega,\Omega')=\frac{1}{4\pi|\bm x_{\textsc{b}}-\bm x_{\textsc{a}}|} \\
\nn&\times
\int\d \omega\tilde\chi_\textsc{a}(-\omega+\Omega)
  \tilde\chi_\textsc{b}(\omega-\Omega')e^{\ii \omega |\bm x_{\textsc{b}}-\bm x_{\textsc{a}}|}.
\end{align}

Mimicking the calculation made before in this section, we further assume that the switching functions are exponentially localized with a time scale ${T}$ and identical up to a shift in time, that is
\begin{align}
    \chi_\textsc{a,b}(s)=\xi\left(\frac{s-t_\textsc{a,b}}{{T}}\right).
\end{align}
These assumptions imply that
\begin{align}
 & I(\Omega,\Omega')=\frac{{T}^2e^{\ii t_\textsc{a}\Omega}e^{\ii t_\textsc{b}\Omega'}}{4\pi|\bm x_{\textsc{b}}-\bm x_{\textsc{a}}|} \\
\nn&\times
\int\d \omega\tilde\xi(-{T}\omega+{T}\Omega)
  \tilde\xi({T}\omega-{T}\Omega')e^{-\ii \omega (t_{\textsc{b}}-t_\textsc{a}-|\bm x_{\textsc{b}}-\bm x_{\textsc{a}}|)}\\
  \nn& =\frac{{T} e^{\ii t_\textsc{a}\Omega}e^{\ii t_\textsc{b}\Omega'}}{4\pi|\bm x_{\textsc{b}}-\bm x_{\textsc{a}}|} \\
\nn&\times
\int\d \upkappa\tilde\xi(-\upkappa+{T}\Omega)
  \tilde\xi(\upkappa-{T}\Omega')e^{-\ii \frac{\upkappa}{{T}} (t_{\textsc{b}}-t_\textsc{a}-|\bm x_{\textsc{b}}-\bm x_{\textsc{a}}|)}
\end{align}
where we have defined the dimensionless integration variable $\upkappa={T}\omega$. By assuming exponential localization we have assumed that the function $\tilde \xi$ admits an analytic extension into the strip $|\Im(\upkappa)|\leq 1$ whose sections $\Im (\upkappa)=$ constant are $L^2$, as explained in Section~\ref{betweentwolevels}. Therefore, we can shift the contour of integration by an imaginary amount $\ii$:
\begin{align}\label{explota}
 & I(\Omega,\Omega')=\frac{{T} e^{\ii t_\textsc{a}\Omega}e^{\ii t_\textsc{b}\Omega'}e^{ \frac{1}{{T}} (t_{\textsc{b}}-t_\textsc{a}-|\bm x_{\textsc{b}}-\bm x_{\textsc{a}}|)}}{4\pi|\bm x_{\textsc{b}}-\bm x_{\textsc{a}}|} \\
\nn&\times
\int\d \upkappa\tilde\xi(-\upkappa-\ii+{T}\Omega)
  \tilde\xi(\upkappa+\ii-{T}\Omega')e^{-\ii \frac{\upkappa}{{T}} (t_{\textsc{b}}-t_\textsc{a}-|\bm x_{\textsc{b}}-\bm x_{\textsc{a}}|)}
\end{align}
and taking the complex modulus, we can bound the value of the integral by
\begin{align}
 & |I(\Omega,\Omega')|\leq\frac{{T} e^{ \frac{1}{{T}} (t_{\textsc{b}}-t_\textsc{a}-|\bm x_{\textsc{b}}-\bm x_{\textsc{a}}|)}}{4\pi|\bm x_{\textsc{b}}-\bm x_{\textsc{a}}|} \\
\nn&\times
\int\d \upkappa|\tilde\xi(-\upkappa-\ii+{T}\Omega)
  ||\tilde\xi(\upkappa+\ii-{T}\Omega')|\\
  \nn &\leq \frac{C{T} e^{ \frac{1}{{T}} (t_{\textsc{b}}-t_\textsc{a}-|\bm x_{\textsc{b}}-\bm x_{\textsc{a}}|)}}{4\pi|\bm x_{\textsc{b}}-\bm x_{\textsc{a}}|}
\end{align}
where $C$ is a constant independent of $\Omega,\Omega'$, which is determined by the specific form factor of the dimensionless switching function $\xi$.

We conclude that the maximum value of the signalling estimator is bounded by
 \begin{align}\label{pointlikebound}
       \mathcal{S}_{\text{max}}\leq  \frac{4C^2{T}^2 e^{ \frac{2}{{T}} (t_{\textsc{b}}-t_\textsc{a}-|\bm x_{\textsc{b}}-\bm x_{\textsc{a}}|)}}{8\pi^2|\bm x_{\textsc{b}}-\bm x_{\textsc{a}}|^2}.
\end{align}
From this example, we learn that the signalling between pointlike detectors can still be bounded exponentially as $|\bm x_{\textsc{b}}-\bm x_{\textsc{a}}|$ goes to infinity, with a scale given by the switching scale ${T}$. Recall that as $|\bm x_{\textsc{b}}-\bm x_{\textsc{a}}|$ grows for a fixed value of  $t_{\textsc{b}}-t_\textsc{a}$, we are estimating the amount of information that a detector A can signal to a detector B that is mostly in spacelike separation and only communicates with A through the switching `tails'. Therefore, this shows that  this ``apparent'' faster-than-light signalling in this scenario is exponentially bounded.

Similarly, the signalling estimator is exponentially bounded as $t_{\textsc{b}}-t_\textsc{a}$ goes to minus infinity with a scale given by ${T}$, thus bounding the apparent retrocausal effects present when a detector A is exponentially localized in the causal future of detector B.

We conclude that the bounds presented for extended detectors are in general too conservative as one approaches the pointlike limit, as far as the detectors are separated in space. We conclude that the  pointlike limit does not necessarily present problems with any apparent faster-than-light signalling or retrocausation.

However, we also observe, e.g. from expression \eqref{explota}, that in the pointlike limit the (exact) expression for our signalling estimator diverges as $|\bm x_{\textsc{b}}-\bm x_{\textsc{a}}|^{-2}$ as the detectors get closer in space, regardless of the localization in time of the detectors. Of course, if the switchings have disjoint supports there is no problem, since in the massless case in four spacetime dimensions the retarded propagator is strictly localized on (the boundary of) the future lightcone, and since the switchings are strictly timelike, the  signalling estimator is identically zero as the propagator evaluates to strictly zero.

However, if the detectors' worldlines are identical (one follows the other) and the switching functions are not compactly supported---and therefore they overlap--- the signalling estimator is divergent, or at least cannot be calculated/estimated within perturbation theory. The physical picture is the following: consider two detectors A and B, that are approximately located at the same position and are both such that they interact with the quantum field for much longer than both the light-crossing time of their characteristic size or their separation. If the detectors are switched on in such a way that they are both interacting with the field at the same time, no matter how weak this overlap is, then the pointlike limit is not well defined. 

The ill-definiteness of timelike signalling in the case of detectors that interact through the same worldline in our example implies that one has three options: a) Either one regularizes the expressions through some sort of renormalization procedure, b) one needs to account for the finite spatial extension of the detector at the level of the interaction Hamiltonian, c) One has to restrict the analysis to disjoint switching functions. There are good reasons to believe that the same problems will appear in the case of other pointlike detector models in general spacetimes. Indeed, the reason why the signalling estimator diverges is that it involves the coincidence limit of the pullback of the retarded propagator into a wordline. Although we will not discuss this here in much detail, it is known that, while the pullback into a one-dimensional timelike curve of some distributions like the Wightman function is well-defined as a distribution over switching functions, the pullback of the retarded propagator is ill-defined. This is a fairly general fact, and although the character of the divergence may depend on the characteristics of the spacetime background, there is no Klein-Gordon field whose propagator admits a well-defined pointlike limit. Given this, it is easy to argue that the rationale applied to the UDW model applies to more complex models describing pointlike interactions. In realistic scenarios, pointlike detectors are, after all, idealizations used to model finite-sized systems so a regularization via a spatial profile would be a good cure for these issues whenever they appear. 

\subsection{More general trajectories}
In the analysis we have performed we have made assumptions about the detectors' trajectories in spacetime. In particular, we assumed that the spacetime smearings factorize in a particular coordinate system for both detectors. This is certainly the case if both detectors do not move relativistically with respect to each other in this reference frame. However, this assumption was not especially relevant at any point in the calculation. This is because, at leading order, the signalling estimator does not explicitly depend on which reference frame each of the detectors is at rest. Indeed, the signalling estimator is given by the formula
\begin{align}
    \mathcal{S}=\text{Var}\;G_{\textsc{r}}[ \hat J_\textsc{b},\braket{\hat J_\textsc{a}}],
\end{align}
which is a frame-independent covariant expression, and more importantly, it does not depend on whether the detectors' smearings and switchings factorize in the same reference frame.

Let us further clarify this point. If one were taking into account higher-order contributions to signalling, these may very well depend on the choice of foliation given the non-microcausal character of the interaction when it is not pointlike (see~\cite{2020broken} for details). However, as far as we restrict our analysis to leading order there is nothing preventing us from defining each of the detectors' spacetime smearings to factorize in different reference frames. Indeed, the joint dynamics of the detectors and the field can be defined with respect to different reference frames, and while technically that may give different predictions, only a subset of these reference frames will, depending on the situation, capture the essence of real physical interactions (this is thoroughly discussed in~\cite{2020broken}). However, and remarkably, the signalling between pairs of detectors will remain unaffected by this choice, so as far as the approximation of Fermi-Walker rigidity~\cite{PhysRevD.101.045017} holds, it makes sense to define each of the detectors in their own reference frames as internally non-relativistic, the predictions of the model will therefore be reliable concerning signalling.

Let us analyze the case where the detectors are identical in the respective reference frames, but the reference frame of detector A is related not only by a simple spacetime translation but also by a fixed Lorentz transformation, implying that it follows a relativistic trajectory from the point of view of detector B. Thus we set the smearing of detector B to be of the form\footnote{The notation follows the definition in~\eqref{lambdaboost}.}
\begin{align}
 \Lambda_{\textsc{b}}^+(\mathsf{x})=\frac{{T}}{|L|}\Xi(L^{-1}\mathsf{x})e^{\ii\Omega \mathsf{u} \mathsf{x}}
\end{align}
and for detector A
\begin{align}
 \Lambda_{\textsc{a}}^+(\mathsf{x})=\frac{{T}}{|L|}\Xi(L^{-1}\mathcal{L}(\mathsf{x}-\mathsf{z}) )e^{\ii\Omega  \mathsf{u} \mathcal{L}(\mathsf{x}-\mathsf{z}) }
\end{align}
where now $\mathcal{L}$ is a matrix representing a (proper orthochronus) Lorentz transformation. Note that we are expressing both smearings in terms of the coordinates for which the switching and smearing of detector B (the receiver) factorize in space and time.

The expression for the integral~\eqref{Iboost} can be again written in terms of Fourier transforms, exploiting the fact that the retarded propagator is translationally invariant:
\begin{align}
&I(\Omega,\Omega')\\
 \nn &={T}^2\!\!\int \d ^{n+1}\mathsf k \, \tilde{G}_\textsc{r}(\mathsf{k}) \tilde{\Xi}^*(L\mathsf{k}+\Omega L\mathsf{u})\tilde{\Xi}(L\mathcal{L}\mathsf{k}+\Omega' L\mathsf{u})e^{-\ii\mathsf{k} \mathsf{z}},
\end{align}
where we have used the fact that
\begin{align}
 &\nn\tilde{\Lambda}_{\textsc{a}}^+(\mathsf{k})\\
 &\nn=\frac{{T}}{{2\pi}^{\frac{n+1}{2}}|L|}\int\d ^{n+1}\mathsf x \Xi(L^{-1}\mathcal{L}(\mathsf{x}-\mathsf{z}) )e^{\ii\Omega  \mathsf{u} \mathcal{L}(\mathsf{x}-\mathsf{z})}e^{-\ii\mathsf{k}\mathsf{x}}\\
  &\nn=\frac{{T}|\mathcal{L}|e^{-\ii\mathsf{k} \mathsf{z}}}{{2\pi}^{\frac{n+1}{2}}}\int\d ^{n+1}\mathsf y \Xi(\mathsf{x})e^{\ii\Omega  \mathsf{u}L \mathsf{y}}e^{-\ii\mathsf{k}\mathcal{L}^{-1}L\mathsf{y}}\\
  &=\tilde{\Xi}(L\mathcal{L}\mathsf{k}+\Omega' L\mathsf{u})e^{-\ii\mathsf{k} \mathsf{z}},
\end{align}
which in turn is true because for proper Lorentz transformations $|\mathcal{L}|=1$ and 
\begin{align}
    \mathsf{x} \mathcal{L}^{-1}\mathsf{y}=(\mathcal{L}\mathsf{x})\mathsf{y}
\end{align}
for all spacetime vectors $\mathsf{x}$ and $\mathsf{y}$.
Now, remember that the dimensionless function $\tilde{\Xi}(\upkappa)$ admits an analytic extension for complex values of its argument fulfilling that the Euclidean norm
$\norm{\text{Im}\upkappa}_e\leq1$, which means that we can shift the contour of integration in the complex plane by a vector $\upeta$
\begin{align}\label{intfourier3eta}
&I(\Omega,\Omega')\\
 \nn &={T}^2e^{-\upeta \mathsf{z}}\!\!\int \d ^{n+1}\mathsf k \, \tilde{G}_\textsc{r} (\mathsf{k}-\ii\upeta)\\
 &\nn\times\tilde{\Xi}^*(L\mathsf{k}+\Omega L\mathsf{u}-\ii L\upeta)\tilde{\Xi}(L\mathcal{L}\mathsf{k}+\Omega' L\mathsf{u}-\ii L\mathcal{L}\upeta)e^{-\ii\mathsf{k} \mathsf{z}},
\end{align}
for $\upeta$ future-directed timelike, with $\norm{L\upeta}_e\leq1$ {\it and} $\norm{L\mathcal{L}\upeta}_e\leq1$.

We again find that by bounding the expression in modulus and restricting our analysis to the massless case, we can bound the maximum signalling by
\begin{align}
\mathcal{S}_{\text{max}}\leq\frac{2C^2{T}^2 e^{-2\upeta\mathsf{z}}}{R^{2n}\upeta^4}.
\end{align}
but in this case, $\upeta$ has to fulfill the extra condition of $\norm{L\mathcal{L}\upeta}_e\leq1$. In Section~\ref{betweentwolevels}, we studied whether apparent retrocausal effects were important by setting $\upeta=\frac{\mathsf{u}}{T}$, and we learned that  apparent retrocausal signalling decayed exponentially in the time lapse between the detectors with a scale given by the typical interaction time $T$. In the case in which detector A follows a trajectory with constant velocity $\bm{v}$ we will not be able to choose  $\upeta=\frac{\mathsf{u}}{T}$ since for that choice,
\begin{align}
  \nn&\norm{L\mathcal{L}\upeta}_e=\frac{1}{T}\norm{L\mathcal{L}\mathsf{u}}_e=\frac{\gamma}{T}\sqrt{T^2+R^2|\bm{v}|^2}\\
  &=\sqrt{\frac{1+(R/T)^2|\bm{v}|^2}{1-|\bm{v}|^2}}>1.
\end{align}
However, if we fix $\mathcal{L}\upeta$ to have unit Euclidean norm instead, namely
\begin{align}
    \upeta=\sqrt{\frac{1-|\bm{v}|^2}{1+(R/T)^2|\bm{v}|^2}}\frac{\mathsf{u}}{T},
\end{align}
then $\norm{L\mathcal{L}\upeta}_e=1$ and
\begin{align}
    \norm{L\upeta}_e=\sqrt{\frac{1-|\bm{v}|^2}{1+(R/T)^2|\bm{v}|^2}}<1,
\end{align}
so all conditions are fulfilled. In this case, considering $\mathsf{z}$ to be past timelike, the signalling estimator is bounded by
\begin{align}\label{retroboundvel}
       \mathcal{S}_{\text{max}}\leq\left(\frac{1+(R/T)^2|\bm{v}|^2}{1-|\bm{v}|^2}\right)^2\frac{4{T}^6C^2}{R^{2n}}e^{-2\frac{|z^0|}{T'}},
\end{align}
with a scale given by 
\begin{align}
    T'=T \sqrt{\frac{1+(R/T)^2 |\bm{v}|^2}{1-|\bm{v}|^2}}.
\end{align}
This scale depends on the typical interaction time of the detectors $T$ as well as on the typical size of their spatial profile. This is an intuitive result, given that the overlap of the detectors is not only regulated by the switching scale but also by the spatial smearing scale for detectors that do not factorize in the same reference frame. It is interesting to note that when  $(R/T)|\bm{v}|\ll1$ (i.e., in the pointlike limit for a fixed $\bm{v}$) the decay scale is given by
\begin{align}
    T'\sim \gamma T,
\end{align}
independent of $R$ (where $\gamma$ is the Lorentz factor associated to the boost relating the two detectors' trajectories). Conversely, for $(R/T)|\bm{v}|\gg1$,  the decay scale is given by
\begin{align}
    T'\sim \gamma|\bm{v}| R,
\end{align}
independent of $T$. Basically, we see how the intuition from studying comoving detectors actually carries to more general trajectories relatively straightforwardly.

\section{Conclusions}\label{sec:conclusions}

In the context of particle detector models, in this work we have analyzed the effect of the infinite `tails' of non-compact detector-field interactions on signalling. We analyzed the apparent signalling and retrocausation that the tails of non-compact detectors can enable when two detectors couple to the same field. Concretely, we have derived a signalling estimator inspired by optimal parameter estimation. By analysing the joint dynamics of pairs of detectors coupled to the field we show explicitly how the signalling between the detectors is governed by the field's Green's function and the internal parameters of the detectors and their interaction with the field. In particular, we extended previous results~\cite{PhysRevD.92.104019} showing that the Fisher information related with the estimation of parameters of A by local measurements on B can be bounded by the variance of an operator built out of the field's Green's function and specific degrees of freedom of the detectors.

Armed with this knowledge, we found that the signalling between non-compactly supported detectors enabled by the tails of the smearing and switching functions depends on the internal dynamics of the detectors. This means that in scenarios with two non-compactly supported detectors (in space or time), one may need to assess case by case whether the detectors are  effectively spacelike separated or not based on the specific parameters characterizing the dynamics of the detectors. However,  at least in the case of flat spacetime, we have shown that signalling always obeys the intuition that if the localization of the detectors decays at least exponentially,  two detectors with infinite tails can indeed be made to be effectively spacelike separated (or more in general, causally orderable, so that the results in~\cite{PhysRevD.103.085002} apply at the level of perturbation theory). This provides further evidence that with smeared particle detectors (even with non compact smearing and switching) one can still make statements about the causality of the model or model scenarios where two detectors are (to all intent and purposes) causally disconnected.

Most of the expressions that we derived for signaling are general enough to be applied in  general globally hyperbolic spacetimes. It would be interesting to explicitly study the decay properties of our signaling estimator in some particular cases e.g. FRW spacetimes, or in the presence of horizons, where smeared detector models may introduce `information leakage' between causally disconnected parts of  spacetime.

\acknowledgments

  EMM acknowledges support through the Discovery Grant Program of the Natural Sciences and Engineering Research Council of Canada (NSERC). EMM also acknowledges support of his Ontario Early Researcher award.
MP acknowledges support of the 2022 Constantine and Patricia Mavroyannis Scholarship Award by the AHEPA Foundation.

\appendix 

\section{Derivation of the operator \texorpdfstring{$\hat{\Sigma}$}{Sigma}} \label{app:sigma}

Consider the case of two general detectors, A and B, which interact with the field according to the interaction Hamiltonian
\begin{align}
   \sum_{\nu=\text{A,B}} \hat H_{\nu}(\tau)=\sum_{\nu=\text{A,B}}\int_{\mathcal{E}(\tau)}\!\!\!\!\!\d\mathcal{E}\;   \hat h_{\nu}(\mathsf x).
\end{align}
where in this case the corresponding Hamiltonian densities will be given by 
\begin{align}
    &\hat h_\textsc{a}(\mathsf x)=\lambda_\textsc{a} \hat J_\textsc{a}(\mathsf x)\otimes\openone_\textsc{b}\otimes\hat \phi(\mathsf x) 
\end{align}
and
\begin{align}
    &\hat h_\textsc{b}(\mathsf x)=\lambda_\textsc{b}  \openone_\textsc{a}\otimes\hat J_\textsc{b}(\mathsf x)\otimes\hat \phi(\mathsf x). 
\end{align}
  The joint evolution in the interaction picture of the detectors and the field can be described as a unitary operator acting over the joint initial state of the field-detectors system $\hat\rho_{\text{initial}}$. Then the state in the asymptotic future will be given by the transformation
\begin{align}
    \hat\rho_{\text{final}}=\hat U_\textsc{a+b}\;\hat\rho_{\text{initial}}\;{\hat U}^{\dagger}_\textsc{a+b}. 
\end{align}
 The unitary implementing the time evolution can be formally written in terms of the Dyson series
\begin{align}
  &\nn\hat U_\textsc{a+b} =\sum_{n}\frac{(-\ii)^n}{n!} \int_{-\infty}^{\infty}\dots\int_{-\infty}^{\infty} \d \tau^{n}\\
&  \times \mathcal{T}\left(\hat H_\textsc{a}(\tau_1)+\hat H_\textsc{b}(\tau_1)\right)\dots\left(\hat H_\textsc{a}(\tau_n)+\hat H_\textsc{b}(\tau_n)\right).
\end{align}
This means that we can rewrite~\eqref{19} as 
\begin{align}
   &\nn\hat U_\textsc{a+b}\;\hat\rho_{\text{initial}}\;{\hat U}^{\dagger}_\textsc{a+b}=\sum_{n}\frac{(-\ii)^n}{n!} \int \d \tau^{n}\\
&  \times \mathcal{T}\left[\hat H(\tau_n),\dots [\hat H(\tau_1),\hat\rho_{\text{initial}}]\dots\right].
\end{align}
If the couplings are weak, we can truncate the series at next to leading order 
\begin{align}
   &\nn\hat U_\textsc{a+b}\;\hat\rho_{\text{initial}}\;{\hat U}^{\dagger}_\textsc{a+b}=\hat\rho_{\text{initial}}\\
 &\nn  -\ii\int_{-\infty}^{\infty} \d \tau
[\hat H_\textsc{a}(\tau)+\hat H_\textsc{b}(\tau),\hat\rho_{\text{initial}}]\\
& \nn -\frac{1}{2}\int_{-\infty}^{\infty}\int_{-\infty}^{\infty} \d \tau\d \tau'\mathcal{T} \left[\hat H_\textsc{a}(\tau)+\hat H_\textsc{b}(\tau),\right.\\
&  \left.[\hat H_\textsc{a}(\tau')+\hat H_\textsc{b}(\tau'),\hat\rho_{\text{initial}}]\right]+\mathcal{O}(\lambda^3).
\end{align}
where the time-ordering operator is defined as follows
\begin{align}
   & \mathcal{T}\hat A({t})\hat B({t}')\\
    &\nonumber\coloneqq\theta({t}-{t}')\hat A({t})\hat B({t}')+\theta({t}'-{t})\hat B({t}')\hat A({t})
\end{align}
for two time-dependent operators $\hat A({t})$ and $\hat B({t})$. The local statistics of the detector B will be given by the partial trace
\begin{align}
    \hat\rho_\textsc{b}=\tr_{\textsc{a},\phi}(\hat U_\textsc{a+b}\;\hat\rho_{\text{initial}}\;{\hat U}^{\dagger}_\textsc{a+b}).
\end{align}
and the signalling term can be defined as 
\begin{align}
  & \nn \hat\rho^{(2)}_{B,\text{sign}}\\
  &=  \frac{\partial^2}{\partial{\lambda_\textsc{a}}\partial{\lambda_\textsc{b}}}\tr_{\textsc{a},\phi}(\hat U_\textsc{a+b}\;\hat\rho_{\text{initial}}\;{\hat U}^{\dagger}_\textsc{a+b})|_{\lambda_\textsc{a}=\lambda_\textsc{b}=0} . 
  \end{align}
Note that, given any operator $\hat O$, it follows that 
\begin{align}
    \tr_{\textsc{a},\phi}([\hat H_\textsc{a}(\tau),\hat O])=0 \label{x}
\end{align}
since $\hat H_\textsc{a}(\tau)$ only depends on operators of detector A and the field, and thereby can be permuted within the partial trace.
This allows us to disregard multiple terms in \eqref{x}, thereby leading to
\begin{align}
\nn   &\hat\rho_\textsc{b}=\tr_{\textsc{a},\phi}(\hat\rho_{\text{initial}})  -\ii\int_{-\infty}^{\infty} \d \tau
\tr_{\textsc{a},\phi}[\hat H_\textsc{b}(\tau),\hat\rho_{\text{initial}}]\\
& \nn -\frac{1}{2}\int_{-\infty}^{\infty}\int_{-\infty}^{\infty} \d \tau\d \tau' \tr_{\textsc{a},\phi}\mathcal{T}\left[\hat H_\textsc{b}(\tau),[\hat H_\textsc{b}(\tau'),\hat\rho_{\text{initial}}]\right]\\
& \nn -\frac{1}{2}\int_{-\infty}^{\infty}\int_{-\infty}^{\infty} \d \tau\d \tau' \tr_{\textsc{a},\phi}\mathcal{T}\left[\hat H_\textsc{b}(\tau),[\hat H_\textsc{a}(\tau'),\hat\rho_{\text{initial}}]\right]\\
&+\mathcal{O}(\lambda^3),
\end{align}
or, more conveniently, we can use the Jacobi identity in the last commutator and again the cyclic property of the partial trace acting over $\hat H_\textsc{a}(\tau)$:
\begin{align}
\nn  & \hat\rho_\textsc{b}=\tr_{\textsc{a},\phi}\hat\rho_{\text{initial}}  -\ii\int_{-\infty}^{\infty} \d \tau
\tr_{\textsc{a},\phi}[\hat H_\textsc{b}(\tau),\hat\rho_{\text{initial}}]\\
& \nn -\frac{1}{2}\int_{-\infty}^{\infty}\int_{-\infty}^{\infty} \d \tau\d \tau' \tr_{\textsc{a},\phi}\mathcal{T}\left[\hat H_\textsc{b}(\tau),[\hat H_\textsc{b}(\tau'),\hat\rho_{\text{initial}}]\right]\\
& \nn -\frac{1}{2}\int_{-\infty}^{\infty}\int_{-\infty}^{\infty} \d \tau\d \tau' \tr_{\textsc{a},\phi}\left[\mathcal{T}[\hat H_\textsc{b}(\tau),\hat H_\textsc{a}(\tau')],\hat\rho_{\text{initial}}\right]\\
&+\mathcal{O}(\lambda^3).
\end{align}
Note that the first three terms do not contribute to the signalling estimator \eqref{estimator}, since they do not depend on $\lambda_\textsc{a}$. Therefore the signalling term will be given by
\begin{align}
  & \nn \hat\rho^{(2)}_{B,\text{sign}}\\
  \nn&=-\frac{1}{2}\int_{-\infty}^{\infty}\int_{-\infty}^{\infty} \d \tau\d \tau' \int_{\mathcal{E}(\tau)}\!\!\!\!\!\d\mathcal{E}\int_{\mathcal{E}(\tau')}\!\!\!\!\!\d\mathcal{E}'\\
 & \times \tr_{\textsc{a},\phi}\left[\hat J_\textsc{a}(\mathsf x')\otimes\hat J_\textsc{b}(\mathsf x)\otimes\mathcal{T}[\hat \phi(\mathsf x),\hat \phi(\mathsf x')],\hat\rho_{\text{initial}}\right].
  \end{align}
Now, consider that the state is initially uncorrelated, i.e. $\hat\rho_{\text{initial}}=\hat\rho_\textsc{a} \otimes \hat\rho_\textsc{b}\otimes \hat\rho_{\phi}$. This gives the following compact expression
\begin{align}
    \hat\rho^{(2)}_{B,\text{sign}}=-\ii[\hat{\Sigma},\hat\rho_\textsc{b}] \label{sigma1}
\end{align}
where we have defined the operator
\begin{align}
    \nn&\hat{\Sigma}=\int\int\d V\d V' \braket{\hat J_\textsc{a}( \mathsf x')}G_\textsc{r}(  \mathsf x,\mathsf x')\hat J_\textsc{b}( \mathsf x)\\
    &=\int\d V G_{\textsc{r}}[\braket{\hat J_\textsc{a}}](  \mathsf x)\hat J_\textsc{b}( \mathsf x). \label{sigma2}
\end{align}
Here $G_\textsc{r}(\mathsf x,\mathsf x')$ is the retarded Green function 
\begin{align}
   G_\textsc{r}(  \mathsf x,\mathsf x')=-\ii \theta(\tau(\mathsf{x})-\tau(\mathsf{x'}))\braket{[\hat\phi(\mathsf{x}),\hat\phi(\mathsf{x}')]}\label{sigma3},
\end{align}
whereas $\d V$ denotes the element of volume with respect to the background metric
\begin{align}
    \d V=\d \mathsf{x}^{n+1}\sqrt{|g|},
\end{align}
and where $\sqrt{|g|}$ is the determinant of the metric. Note that we have used the fact that
\begin{align}
    &\nn\int \d \tau \int\d\mathcal{E}(\tau) =\int\d \mathsf{x}^{n+1}\sqrt{|g|}  \int\d\tau\delta(\tau(\mathsf{x})-\tau)\\
    &=\int\d\mathsf{x}^{n+1}\sqrt{|g|}=\int\d V.
\end{align}

\section{Frequency independent bound}\label{app:INTBound}

The bounds \eqref{b1}, \eqref{b2} on the propagator give the following maximum values of the integral \eqref{intfourierbound}:
\begin{align}\label{intfourierbound2}
 \nn& |I(\Omega,\Omega')|
\leq \frac{1}{(2\pi)^{(n+1)}}\frac{e^{-\upeta\mathsf{z}}}{-\upeta^2}\\
&\times\int \d ^{n+1}\mathsf k \, |\tilde{\Xi}^*(L\mathsf{k}+\Omega L\mathsf{u}-\ii L\upeta)\tilde{\Xi}(L\mathsf{k}+\Omega' L\mathsf{u}-\ii L\upeta)|
\end{align}
for fields with small enough mass (in particular massless fields), and
\begin{align}\label{intfourierbound3}
 \nn& |I(\Omega,\Omega')|
\leq \frac{1}{(2\pi)^{(n+1)}}\frac{e^{-\upeta\mathsf{z}}}{2m\sqrt{-\upeta^2}}\\
&\times\int \d ^{n+1}\mathsf k \, |\tilde{\Xi}^*(L\mathsf{k}+\Omega L\mathsf{u}-\ii L\upeta)\tilde{\Xi}(L\mathsf{k}+\Omega' L\mathsf{u}-\ii L\upeta)|
\end{align}
for very massive fields.

Now, the factor under the integral sign is of the form
\begin{align}
    \int |f^*g|
\end{align}
for two $L^2$ functions. We can use the Cauchy-Schwartz inequality:
\begin{align}
      \left(\int |f^*g|\right)^2\leq \int|f|^2\int|g|^2
\end{align}
to conclude that
\begin{align}
    \nn&\int \d^n\mathsf k \, |\tilde{\Xi}^*(L\mathsf{k}+\Omega L\mathsf{u}-\ii L\upeta)\tilde{\Xi}(L\mathsf{k}+\Omega' L\mathsf{u}-\ii L\upeta)|\\
    \nn&\leq\norm{\tilde\Xi(L\cdot+\Omega L\mathsf{u}+\ii L\upeta)}_2\norm{\tilde\Xi(L\cdot+\Omega' L\mathsf{u}-\ii L\upeta)}_2\\
    \nn&=\norm{\tilde\Xi(L\cdot+\ii L\upeta)}_2\norm{\tilde\Xi(L\cdot-\ii L\upeta)}_2\\
    &\nn=\frac{1}{|L|}\norm{\tilde\Xi(\cdot+\ii L\upeta)}_2\norm{\tilde\Xi(\cdot-\ii L\upeta)}_2\\
    &=\frac{1}{|L|}\norm{\Xi e^{-L\upeta \cdot}}_2\norm{\Xi e^{L\upeta \cdot}}_2.
\end{align}

In addition to the Cauchy-Schwartz inequality, in this derivation we have used the following property of the $L^2$ norm
\begin{align}
    \norm{f(L\cdot+a)}_2=\frac{1}{\sqrt{|L|}} \norm{f}_2,
\end{align}
and the fact that
\begin{align}
    \norm{\tilde f(\cdot-\ii L\upeta)}_2= \norm{fe^{L\upeta \cdot}}_2,
\end{align}

Altogether, we have derived a frequency-independent bound:
\begin{align}\label{intfourierbound4}
  |I(\Omega,\Omega')|
\leq \frac{1}{(2\pi)^{(n+1)}}\frac{e^{-\upeta\mathsf{z}}}{|L|(-\upeta^2)}\norm{\Xi e^{-L\upeta \cdot}}_2\norm{\Xi e^{L\upeta \cdot}}_2
\end{align}
for fields with low mass and
\begin{align}\label{intfourierbound5}
 |I(\Omega,\Omega')|
\leq\frac{1}{(2\pi)^{(n+1)}} \frac{e^{-\upeta\mathsf{z}}}{|L|2m\sqrt{-\upeta^2}}\norm{\Xi e^{-L\upeta \cdot}}_2\norm{\Xi e^{L\upeta \cdot}}_2
\end{align}
for fields with high mass.




\section{Signaling estimator for Gaussian smearings and switchings in 3+1 dimensional Minkowski spacetime} \label{APPENDIX2}

In $3+1$ dimensions the Green's function is
\begin{equation}
    G_{\textsc{r}}(t-t',\bm{x}-\bm{x}')= -\frac{1}{4\pi} \frac{\delta (t-t'-|\bm{x}-\bm{x}'|)}{|\bm{x}-\bm{x}'|}
\end{equation}
We will consider Gaussian smearing and switching functions to evaluate the expression 
\begin{align}
    I(\Omega_{\textsc{a}},\Omega_{\textsc{b}})=  \int \d t \d t' \chi_ {\textsc{b}}(t) e^{\ii \Omega_{\textsc{b}} t } \chi_ {\textsc{a}}(t') e^{-\ii \Omega_{\textsc{a}} t' } \mathcal{C}(t-t')
\end{align}
where
\begin{equation}
    \mathcal{C}(t-t')=  -\frac{1}{4\pi}  \int \d \bm{x} \d \bm{x}' F_{\textsc{b}}(\bm{x}) F_{\textsc{a}}(\bm{x}') \frac{\delta(t-t'- |\bm{x}-\bm{x}'|)}{|\bm{x}-\bm{x}'|} \label{c}
\end{equation}

We consider that B is centered around zero, i.e.,
\begin{equation}
  F_  {\textsc{b}}(\bm x )= \frac{1}{(\sqrt{2\pi}R)^3}e^{-\frac{\bm{x}^2}{2R^2}} \label{smearing}
\end{equation}
and A is centered around $\bm{z}$ with the same width $R$. Let's introduce the following non-orthogonal change of variables: we keep the variable $\bm x$ and we define a new integration variable $\bm{y}=\bm{x}-\bm{x}'$ (the Jacobian is 1). We then first perform the integral with respect to $\bm{x}$ in \eqref{c} which is a convolution of the two Gaussian smearings
\begin{equation}
    \int \d \bm{x}  F_  {\textsc{b}}(\bm x ) F_{\textsc{a}}(\bm x- \bm y)= \frac{1}{(2\sqrt{\pi}R)^3}e^{-\frac{(\bm y + \bm z)^2}{4R^2}}
\end{equation}
Then 
\begin{equation}
   \mathcal{C}(t-t')= -\frac{1}{4\pi} \int \d \bm y e^{-\frac{(\bm y + \bm z)^2}{4R^2}}\frac{\delta(t-t'- |\bm{y}|)}{|\bm{y}|}
\end{equation}
Calling $\tau= t-t'$ and $|\bm z|:=L$ 
\begin{align}
    \mathcal{C}(\tau)= -\frac{e^{-\frac{
    L^2}{4R^2}} }{2} \int_0^{\infty} |\bm y| \d |\bm y| e^{-\frac{
    |\bm y |^2}{4R^2}} \frac{2R^2}{L|\bm y|} &\left( e^{\frac{L|\bm{y}|}{2R^2}} -e^{-\frac{L|\bm{y}|}{2R^2}} \right) \nn \\
    &\times \delta (\tau-|\bm y|)
\end{align}
Altogether,
\begin{align}
    I(\Omega_{\textsc{a}},\Omega_{\textsc{b}})=   \frac{ 4\pi e^{-\frac{
    L^2}{4R^2}}}{(2\sqrt{\pi})^3LR} \int |\bm y| \d |\bm y| &e^{-\frac{
    |\bm y |^2}{4R^2}} \left( e^{\frac{L|\bm{y}|}{2R^2}} -e^{-\frac{L|\bm{y}|}{2R^2}} \right) \nonumber \\
    &\times I_p(|\bm{y}|,\Omega_A, \Omega_B)
\end{align}
where $ I_p(|\bm{y}|,\Omega_A, \Omega_B)$ is the expression of $I$ for two pointlike detectors with identical switchings separated by a distance $|\bm y|$,
\begin{align}
    I_p&(|\bm{y}|,\Omega_A, \Omega_B) \nonumber \\
    &=-\frac{1}{4\pi} \int \d t \d t' \chi_ {\textsc{b}}(t) e^{\ii \Omega_{\textsc{b}} t } \chi_ {\textsc{a}}(t') e^{-\ii \Omega_{\textsc{a}} t' } \nonumber  \frac{\delta(t-t'- |\bm{y}|)}{|\bm{y}|}\\
    &=-\frac{1}{4\pi} \int \d t \d \tau \chi_{\textsc{b}}(t)e^{\ii (\Omega_{\textsc{b}}-\Omega_{\textsc{a}}) t } \chi_{\textsc{a}}(t-\tau)e^{\ii \Omega_{\textsc{a}}\tau }\frac{\delta(\tau- |\bm{y}|)}{|\bm{y}|}\nonumber  \\
    &=-\frac{e^{\ii \Omega_{\textsc{a}}|\bm{y}|}}{4\pi|\bm{y}|} \int \d t \chi_{\textsc{b}}(t)\chi_{\textsc{a}}(t-|\bm{y}|)e^{\ii (\Omega_{\textsc{b}}-\Omega_{\textsc{a}}) t }
\end{align}
Now we evaluate this for Gaussian switching functions
\begin{align}
\chi_{\textsc{b}}(t)= \frac{1}{\sqrt{2\pi}}e^{-\frac{t^2}{2T^2}} \label{switching}
\end{align}
and A displaced by $\Delta$ we have 
\begin{align}
     I_p(|\bm{y}|,\Omega_{\textsc{a}}, \Omega_{\textsc{b}})= -\frac{e^{\ii \Omega_{\textsc{a}}|\bm{y}| }}{8\pi^2 |\bm{y}|}&\int \d t e^{\ii (\Omega_{\textsc{b}}-\Omega_{\textsc{a}})t} \nn \\
     & \times e^{-\frac{t^2}{2T^2}} e^{-\frac{(t-(|\bm{y}|+\Delta))^2}{2T^2}}
\end{align}
which is a Gaussian integral that gives
\begin{align}
   I_p(|\bm{y}|,\Omega_{\textsc{a}}, \Omega_{\textsc{b}})= \frac{T\sqrt{\pi}}{8 \pi^2 |\bm{y}|} & e^{\frac{\ii (\Omega_{\textsc{a}}+ \Omega_{\textsc{b}})|\bm{y}|}{2}}  e^{\frac{i (\Omega_{\textsc{b}}-\Omega_{\textsc{a}})\Delta}{2}} \nn \\
   & \times e^{-\frac{(\Omega_{\textsc{b}}-\Omega_{\textsc{a}} )^2 T^2}{4}} e^{-\frac{(|\bm{y}|+\Delta)^2}{4 T^2} }
\end{align}
Finally, changing variables $u:=|\bm y|/R$ we get
\begin{align}
I(\Omega_A, \Omega_B)= \frac{R}{\sqrt{\pi}L} e^{-\frac{L^2}{4R^2}}\int_0^{\infty} \d u \; u & e^{\frac{-u^2}{4}}(e^{\frac{uL}{2R}}-e^{-\frac{uL}{2R}}) \nn \\
& \times I_p(Ru,\Omega_A, \Omega_B)
\end{align}
where
\begin{align}
  I_p(L,\Omega_A, \Omega_B)=  -\frac{T}{8\pi^{3/2}L} e^{-\frac{T^2(\Omega_B-\Omega_A)^2}{4}} e^{-\frac{(L+\Delta)^2}{4T^2}} \nn \\
  \times e^{\ii (\Omega_B+\Omega_A)L}.
\end{align}

\bibliography{refs.bib}

\end{document}